\documentclass[prl,twocolumn,showpacs,preprintnumbers,amsmath,amssymb,superscriptaddress]{revtex4-2}

\usepackage{graphicx}
\usepackage{dcolumn}
\usepackage{bm}
\usepackage{bbm}
\usepackage{color}
\usepackage{amsfonts}
\usepackage{soul}
\usepackage[colorlinks=true,citecolor=blue,linkcolor=blue,urlcolor=blue]{hyperref}

\begin{document}
\title{Controllable Operations of Edge States in Cross-One-dimensional Topological Chains}
\author{Xian-Liang Lu}
\affiliation{School of Physics, Sun Yat-sen University, Guangzhou 510275, China }
\author{Ze-Liang Xiang}
\email{xiangzliang@mail.sysu.edu.cn}
\affiliation{School of Physics, Sun Yat-sen University, Guangzhou 510275, China }
\date{\today}
\begin{abstract}
Topological edge states are recently attracting intense interest due to their robustness in the presence of disorder and defects. However, most approaches for manipulating such states require global modulations of the system's Hamiltonian. In this work, we develop a method to control edge states using local interactions of a four-node junction between cross-one-dimensional topological atomic chains. These junction interactions can give rise to tunable couplings between the hybridized edge states within different geometric symmetry, allowing us to implement robust quantum state transfer and SWAP gate between the two topological chains, where the edge states are pair-encoded as a single qubit. Moreover, when the atoms are precisely positioned to couple waveguides, the correlated decay caused by the environment enables the anti-symmetric edge states to present subradiant dynamics and thus show extremely long coherence time. These findings open up new possibilities for quantum technologies with topological edge states in the future. 
\end{abstract}
\maketitle

As a fundamental element in topological materials, the edge states predicted by the bulk-boundary correspondence emerge at the interface where the topological invariant changes~\cite{hasan2010colloquium,qi2011topological}. One of the most significant features of topological systems is their robustness against disorder and defects~\cite{halperin1982quantized,niu1984quantised,buttiker1988absence}, which leads to topologically protected energy information flow~\cite{Hafezi2011,perczel2017topological,bandres2018topological}. In addition, the zero-energy edge states~\cite{ryu2002topological} in the one-dimensional~(1D) topological system with chiral Hamiltonian can separate from bulk states with large bandgaps~\cite{asboth2016short,nie2020bandgap}. These unique advantages show great potential in realizing robust quantum state transfer with 1D lattices~\cite{yao2013topologically,Dlaska_2017,lang2017topological,mei2018robust,longhi2019landau,d2020fast,qi2021topological}, as well as performing topological quantum computation in superconducting wire networks~\cite{Alicea2011,sarma2015majorana,karzig2017scalable}, where the unpaired Majorana zero modes are encoded as qubits. Thus, achieving the quantum control of such topological edge states has become a critical issue.

The transfer and operations of multiple protected edge states are vital in robust quantum computation and large-scale quantum information processing. Previous works have developed techniques based on adiabatic protocols to drive single edge state~\cite{mei2018robust,longhi2019landau,d2020fast,qi2021topological,brouzos2020fast,tian2022experimental,Pakkiam2023}, in which the operation time is generally limited by the adiabatic theorem without accelerated strategies~\cite{d2020fast}. The topological charge pump, introduced by Thouless~\cite{thouless1983quantization}, can give rise to a quantized transport of particles in each cyclic evolution~\cite{lohse2016thouless,nakajima2016topological}. In recent studies, non-quantized topological pumping is exploited to transfer the edge states~\cite{gu2017topological,longhi2019topological}, and experimental realizations have been reported in a variety of platforms including photonics~\cite{kraus2012topological,verbin2015topological,Zilberberg2018,Cheng2022}, elastic lattices~\cite{rosa2019edge}, magneto- and electro-mechanical systems~\cite{Grinberg2020,xia2021experimental}, and acoustic structures~\cite{cheng2020experimental,chen2021landau,Chen2021}. However, due to the topological robustness, these approaches demand simultaneous modulations of system parameters to drive the edge states, and they are incapable of building multi-qubit operations with edge states. On the other hand, the quantum gates based on braiding in topological chains also require adiabatic moving of domain walls~\cite{lahtinen2017short,boross2019poor}. Thus, it is still challenging to implement direct quantum control of topological edge states with few-body interactions and in non-adiabatic processes.  All of these give rise to the following question: Can one build tunable interactions between edge states to implement such transport and quantum gates through dynamical evolution?

In this Letter, we present a concise approach to control the topological edge states with few-body local interactions, where we leverage the interactions within a four-node junction formed by two intersecting topological chains of two-level atoms, to implement direct manipulations of edge states. In the topologically nontrivial phase, the paired zero-energy edge states~\cite{delplace2011zak,asboth2016short} emerge at four edges of the system and are effectively controlled by the local junction interactions. When we encode the four edge states as two-qubit states, we find that the $C_4$ and $\mathbb{Z}_2$ geometric symmetries of the system can contribute to different qubit-qubit interactions, enabling us to implement the robust state transfer and a topological SWAP gate via the non-adiabatic process. Moreover, we study the dissipative dynamics of the edge states when each atomic chain is specifically structured to couple individual 1D electromagnetic environment, which leads to the topologically protected super/subradiance of edge states. This allows us to generate and transfer remote entanglement between edge atoms. Our proposal provides a means of manipulating edge states via local interactions and paves the way for robust qubit operations as well as long-time storage.

\begin{figure}[t]
\centering
\includegraphics[width=0.9\linewidth]{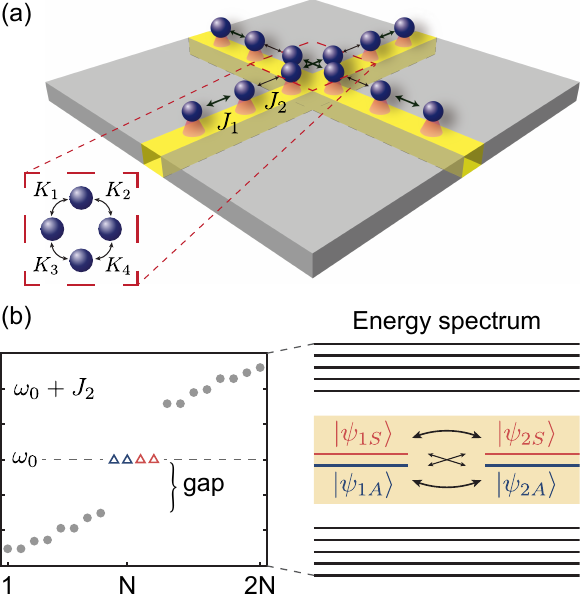}
\caption{(a) Schematic of the compound topological system. Two identical atomic chains with bipartite lattice have staggered hopping amplitudes $J_1$ and $J_2$ respectively, and there are tunable inter-chain hoppings $K_i$ at the central node. Each chain is coupled to an individual waveguide. (b) Eigenenergies of the model (with an $\omega_0$ frequency shift) and the zero-energy manifold composed of the hybridized edge states in the single-excitation subspace. The interactions between edge states are mediated by tunable central network couplings $K_i$.}
\label{Fig1}
\end{figure}

\emph{Model.}---We consider a topological system as depicted in Fig.~\ref{Fig1} (a), where two identical topological chains cross at their mutual center. Each chain consists of $2N$ identical two-level atoms to form a Su-Schrieffer-Heeger (SSH) model~\cite{su1979solitons}, i.e., the atoms are located in two sets of sublattices with alternated hopping rates, which can be described by a tight-binding Hamiltonian. In addition, each chain couples to a waveguide in order to induce nonlocal dissipative couplings between atoms. Such a configuration can be realized with artificial atomic systems, such as superconducting circuits~\cite{xiang2013hybrid,lu2023controllable}. The Hamiltonian of atomic chains reads $H_\mathrm{tot}=H_0+H_{\mathrm{XSSH}}+H_I$, where $H_0=\sum_{l=1,2}\sum_i \omega_a \sigma_{l,i}^{+}\sigma_{l,i}^{-}$ denotes the bare Hamiltonian with the transition frequency $\omega_a$. $H_{\mathrm{XSSH}}$ describes the two chains with no inter-chain hopping as $H_{\mathrm{XSSH}}=\sum_{l=1,2}\sum_i(J_1\sigma_{l,iA}^{+}\sigma_{l,iB}^{-}+J_2\sigma_{l,i+1A}^{+}\sigma_{l,iB}^{-}+\mathrm{H.c.})$,
where $l$ is the chain's index, and $\sigma_{l,iA}^{+(-)}$ and $\sigma_{l,iB}^{+(-)}$ are the raising (lowering) operators for atoms in the corresponding sublattices A and B of the $i$-th unit cell, respectively. The staggered nearest-neighbor hopping amplitudes $J_1$ and $J_2$ represent the inter- and intra-cell hoppings, respectively. In the single-excitation subspace, when $J_1<J_2$ the paired edge states emerge at the two ends of each finite-sized SSH chain, which hybridize from $|\psi_L\rangle$ and $|\psi_R\rangle$ into the symmetric ($|\psi_S\rangle$) and anti-symmetric ($|\psi_A\rangle$) edge states, with energies~\cite{delplace2011zak}
\begin{equation}
\epsilon_S=-\epsilon_A=(-1)^{N+1} \frac{J_2\sinh \lambda}{\sinh(N+1) \lambda},
\end{equation}
where $\lambda$ is given by $\sinh{(N \lambda)}/\sinh{\left[(N+1) \lambda\right]}=J_1/J_2$. In the inset of Fig.~\ref{Fig1} (a), we show the configuration of the central four-node junction. The atomic emitters are coupled through tunable couplers, which contribute to the interactions between the two topological chains,
\begin{equation}
\begin{aligned}
H_I=(&K_1\sigma_{1,s}^{+}+K_2\sigma_{1,s+1}^{+})\sigma_{2,s}^{-}+\mathrm{H.c.}\\
&+(K_3\sigma_{1,s}^{+}+K_4\sigma_{1,s+1}^{+})\sigma_{2,s+1}^{-}+\mathrm{H.c.},
\end{aligned}
\end{equation}
where $s$ and $s+1$ are the location index of the central atoms. Without loss of generality~\cite{lu2023controllable}, the cell number $N$ is assumed to be odd and thus $s=\{(N+1)/2,A\}$, $s+1=\{(N+1)/2,B\}$. There are four well-defined edge states when we consider $H_I$ as a perturbation, which can be decomposed into three parts, corresponding to edge-edge, edge-bulk, and bulk-bulk couplings. It offers different interactions between these hybridized edge states, as illustrated in Fig.~\ref{Fig1} (b), whereas the bulk-edge coupling is suppressed due to the large gap between the edge and bulk states~\cite{nie2020bandgap}. Additionally, this gap also protects the edge states from thermal noise, and thus we focus on the dynamics within the zero-energy manifold of edge states.

\begin{figure}[t]
\centering
\includegraphics[width=0.98\linewidth]{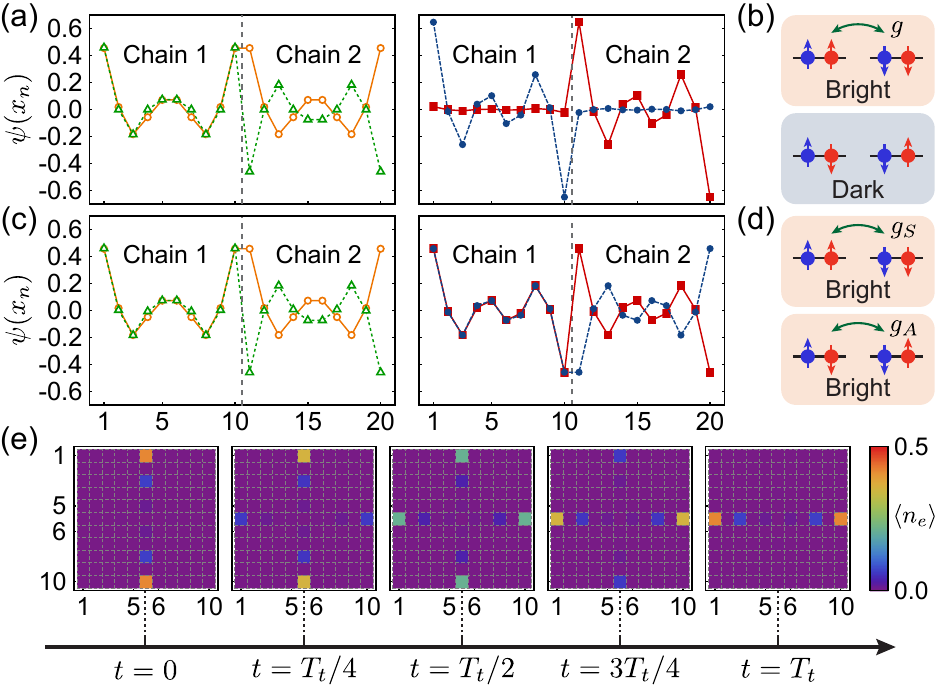}
\caption{Junction-induced tunable interactions between edge states with different geometries. Real-space wave functions of the edge states in (a) and (c) correspond to $C_4$ and $\mathbb{Z}_2$ symmetries, respectively. (b) and (d) show the schematic illustration of tunable spin-exchange interactions between hybridized edge states.  The dark state interacting channel opens when the system breaks into $\mathbb{Z}_2$ symmetry from $C_4$ symmetry. Here we choose the parameters as $J_1=0.4J_2$, with $N=5$ for each chain. The junction interactions are $K_i=0.07J_2$ in (a) and $K_1=K_3=0.07J_2$, $K_2=K_4=0.05J_2$ in (c), respectively. (e) Schematic of the edge-state transfer in a non-adiabatic process. The state transfer is simulated with the initial state $|\psi(0)\rangle=|\psi_\mathrm{1S}\rangle$.}
\label{Fig2}
\end{figure}

\emph{Bright/Dark states and state transfer}---First we consider the simplest case where all the local junction interactions are identical, i.e., $K_i=K$. The system now possesses $C_4$ symmetry and each chain exhibits the inversion symmetry~\cite{lu2023controllable}, leading to degenerate subspaces of odd-parity eigenstates $|\psi_{l,2n-1}\rangle$, with $n=1, \cdots, N$. Here, $|\psi_{l,n}\rangle$ denotes the $n$-th eigenstate of SSH chain in ascending order of energy. Consequently, the symmetric edge states $|\psi_{1S}\rangle,|\psi_{2S}\rangle$ become coupled modes owing to their even parity, as shown in Fig.~\ref{Fig2} (a), and in contrast the anti-symmetric states $|\psi_{1A}\rangle,|\psi_{2A}\rangle$ are uncoupled modes. Using the perturbation theory, we obtain the effective interaction Hamiltonian
\begin{equation}
H_\mathrm{eff}^\mathrm{id}=\widetilde{H}_0+g\left(|\Psi_{1S}\rangle\langle\Psi_{2S}|+\mathrm{H.c.}\right),
\end{equation}
where $\widetilde{H}_0=\sum_{l=1,2}2\epsilon_S\left(|\Psi_{lS}\rangle\langle\Psi_{lS}|-|\Psi_{lA}\rangle\langle\Psi_{lA}|\right)$ is the unperturbed Hamiltonian, with the coupling strength
\begin{equation}
g=2K\eta^2, \quad \eta = \frac{\sinh[(N+1)\lambda/2]}{\sqrt{2 \sum_i \sinh^2[(N+1-i)\lambda]}}.
\end{equation}
We note that the second-order perturbation induced by the residual bulk-edge coupling is considerably weak when $N$ is odd~\cite{lu2023controllable}. In this case, only symmetric edge states have the transition element, which results in an oscillation in their subspace, manifesting a topological state transfer from one chain to another via the local junction in a non-adiabatic process. As shown in Fig.~\ref{Fig2}(b), this leads to the formation of bright and dark states, where we denote the edge states $\{|\psi_{1S}\rangle,|\psi_{1A}\rangle,|\psi_{2A}\rangle,|\psi_{2S}\rangle\}$ as $\{\mid\uparrow\uparrow\rangle,\mid\uparrow\downarrow\rangle,\mid\downarrow\uparrow\rangle,\mid\downarrow\downarrow\rangle\}$, respectively. In, Fig. \ref{Fig2}(e), we show the process of an excitation swap between the bright states of different chains with transfer time $T_t=\pi/2g$, whereas the dark states keep stationary.

As regards the rotationally symmetric configuration $K_1=K_3=K_+$ while $K_2=K_4=K_-$ ($K_+>K_-$), the system remains a $\mathbb{Z}_2$ residual symmetry. The effective Hamiltonian in the zero-energy manifold reads 
\begin{equation}
H_{\rm eff}^{\rm rot}= \widetilde{H}_0 + g_S |\psi_{1S}\rangle\langle\psi_{2S}|+g_A |\psi_{1A}\rangle\langle\psi_{2A}| + \mathrm{H.c.},
\end{equation}
where $g_S = \eta^2(K_+ + K_-)$, $g_A = \eta^2(K_+ - K_-)$.
Fig.~\ref{Fig2}(c) shows that the edge states are shared by two chains and distribute uniformly at the four edges. This distribution, protected by $\mathbb{Z}_2$ symmetry, is independent of the parameters, where the edge states are denoted by $|\psi_S^+\rangle=(1,1,1,1)^\mathrm{T}/2$, $|\psi_S^-\rangle=(1,1,-1,-1)^\mathrm{T}/2$ and $|\psi_A^+\rangle=(1,-1,1,-1)^\mathrm{T}/2$, $|\psi_A^-\rangle=(1,-1,-1,1)^\mathrm{T}/2$ in terms of the basis $\{|\psi_{1L}\rangle,|\psi_{1R}\rangle,|\psi_{2L}\rangle,|\psi_{2R}\rangle\}$. In this configuration, the junction interactions provide an additional interacting channel for the two-qubit exchange process through anti-symmetric edge states, resulting in their conversion from the dark to bright states, as shown in Fig.~\ref{Fig2} (d). These two interacting subspaces correspond to two distinct uncoupled spin-exchange processes. We can perfectly control the switching on/off of the interacting channels and their strength by changing the local junction interactions. In a more general case with broken $\mathbb{Z}_2$ symmetry, more interaction channels become accessible~\cite{lu2023controllable}. 

\emph{Transitions and quantum gates.}---Having a platform with tunable interactions between edge states, we investigate its effective model in the picture of two spins. We solely use the inversion symmetry to reduce its original Hamiltonian in the zero-energy manifold. The effective spin model in the rotationally symmetric scenario has an anisotropic Heisenberg Hamiltonian $H_\mathrm{eff}^\mathrm{rot}=t \sigma_z \otimes \sigma_z + u\sigma_x\otimes\sigma_x-v\sigma_y\otimes\sigma_y$, where $t=2\epsilon_S$, $u=\eta^2 K_+$, $v=\eta^2 K_-$. In the case with $C_4$ symmetry, it reduces to $u=v$, where the rotating wave term $\sigma_1^+\sigma_2^-+\mathrm{H.c.}$ vanishes in the spin-flipping process 
$\mid\uparrow\downarrow\rangle\langle\downarrow\uparrow\mid$ while the counter-rotating wave term $\sigma_1^+\sigma_2^+ +\mathrm{H.c.}$ dominates. When $u\neq v$, the edge states oscillate in two decoupled subspaces $\Pi:\left\{\mid\uparrow\uparrow\rangle,\;\mid\downarrow\downarrow\rangle\right\}$ and $\Sigma:\left\{\mid\uparrow\downarrow\rangle,\;\mid\downarrow\uparrow\rangle\right\}$, respectively, as mentioned above, offering another degree of freedom to manipulate the topological qubit states.

\begin{figure}[t]
\centering
\includegraphics[width=0.98\linewidth]{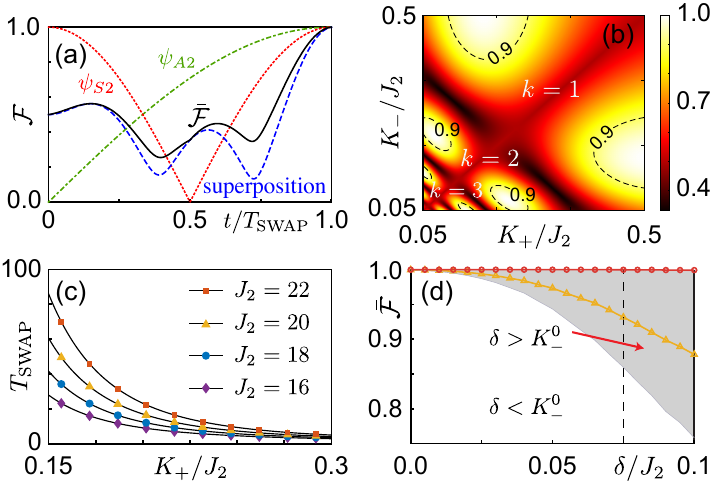}
\caption{(a) The gate fidelity of different initial states. The initial states are $\psi_{S2}$, $\psi_{A2}$, $\left(\psi_{S2}+\psi_{A2}\right)/\sqrt 2$ (blue dashed), respectively. The solid black line denotes average fidelity $\tilde{F}$. (b) The high-fidelity regions with tunable junction interactions $K_i$. (c) Tunable gate time by modifying junction interactions with different values of $J_2$. Here, we set $K_+=3K_-$, and $J_1$ is the optimal value to achieve the highest fidelity. (d) The gate fidelity as a function of the disorder strength $\delta$. The solid lines correspond to the results averaged over $10^3$ disorder instances with (red) and without (yellow) the modulation of $K_i$, where the shaded area indicates a standard deviation for the yellow line (the fluctuation of the red line is less than $0.01$). Here, we take $K_\pm^0$ at $k=2$ regime, $J_1=0.51 J_2$ [except in (c)], and $N =5$.}
\label{Fig3}
\end{figure}

We analyze the coherent dynamics in our system to illustrate the operation of edge-state qubits, specifically focusing on the SWAP gate~\cite{kandel2019coherent}. The time evolution of an arbitrary state is given by
\begin{equation}\label{timeevolution}
\psi(t)=\frac{1}{2}\left(\begin{array}{cccc}
\mathcal{R}_\Pi^{+} & 0 & 0 & \mathcal{R}_\Pi^{-} \\
0 & \mathcal{R}_\Sigma^{+} & \mathcal{R}_\Sigma^{-} & 0 \\
0 & \mathcal{R}_\Sigma^{-} & \mathcal{R}_\Sigma^{+} & 0 \\
\mathcal{R}_\Pi^{-} & 0 & 0 & \mathcal{R}_\Pi^{+}
\end{array}\right) \psi(0)
\end{equation}
where $\mathcal{R}_\Pi^{\pm}=U_S^+\pm U_S^-$ and $\mathcal{R}_\Sigma^{\pm}=U_A^+\pm U_A^-$ are the matrix elements of different subspaces, and the rotating vectors $U_{\nu}(t)=e^{-i E_{\nu}^\pm t / \hbar}$ in the complex plane can contribute to the interference of different components, with energies $E_S^\pm=t\pm(u+v)$ and $E_A^\pm=-t\pm(u-v)$. The oscillation frequencies in the $\Sigma$ and $\Pi$ subspaces are $\Omega_1=2(u-v)$ and $\Omega_2=2(u+v)$, respectively. Therefore, when we adjust the junction interactions such that $K_+=3K_-$, the unitary evolution becomes a SWAP gate at the time
\begin{equation}\label{timeofSWAP}
T_\mathrm{SWAP}=\frac{2n\pi}{2(u+v)}=\frac{m\pi}{2(u-v)}=\frac{2k\pi}{2(t+v)},
\end{equation}
where $n,m,k\in \mathbb{Z}$. As verified by numerical simulation in Fig.~\ref{Fig3}(a), the SWAP gate with different initial states is realized, and the gate fidelity $\bar{\mathcal{F}}>0.999$ after a complete time cycle $T_\mathrm{SWAP}$, where we use the average fidelity $\bar{\mathcal{F}}=\int d\psi_q{\rm Tr}\{\vert\psi_q\rangle\langle\psi_q\vert\rho(t)\}$~\cite{NIELSEN2002249} with $\psi_q$ being the target state.  
The result is calculated by the total Hamiltonian $H_\mathrm{tot}$, which shows good consistency with our analysis based on the effective Hamiltonian $H_\mathrm{eff}^\mathrm{rot}$. Eq.~\eqref{timeofSWAP} determines the sweet point of the junction interactions $(K_-^0,K_+^0)$, with $K_+^0=3K_-^0$, $\epsilon_S=2\eta^2(4k-1) K_-^0$. In Fig.~\ref{Fig3}(b), the high-fidelity regions are separated by different $k$ values, and the diagram reveals the rotationally symmetric structure by exchanging $K_+ \leftrightarrow K_-$. The gate time can also be tuned by modifying the junction interactions $K_\pm$, as depicted in Fig.~\ref{Fig3}(c), which scales as $T_\mathrm{SWAP}\sim K_+^{-N}$ and the fidelity maintains $\bar{\mathcal{F}}>0.99$ within a certain range of $K_+$~\cite{lu2023controllable}. In the SSH chain, the edge states protected by the chiral symmetry show robustness against the imperfections of hopping rates $J_i$, i.e., the off-diagonal disorder. However, the quantum gate operations and state transfer are subject to this disorder since it changes the energy of edge states~\cite{longhi2019landau,lang2017topological}, which results in the breakdown of condition Eq.~\eqref{timeofSWAP}. In our model, the energy levels of edge states are also related to the junction interactions. Thus, by tuning the junction interactions, we can still reconstruct the SWAP gate in Eq.~\eqref{timeevolution} in the presence of the off-diagonal disorder~\cite{lu2023controllable}. In Fig.~\ref{Fig3}(d), we illustrate the gate fidelity with and without the modulation of $K_i$ as a function of the disorder strength $\delta$, where $J_i=J_i+ \delta_i$ and the disorder $\delta_i\in [-\delta,\delta]$ is in uniform distribution. The junction interactions significantly reduce the deviation and fluctuation of the fidelity, and it shows a plateau ($\bar{\mathcal{F}}>0.99$) even when the disorder is larger than the junction interaction $K_-^0$.

\begin{figure}[t]
\centering
\includegraphics[width=0.98\linewidth]{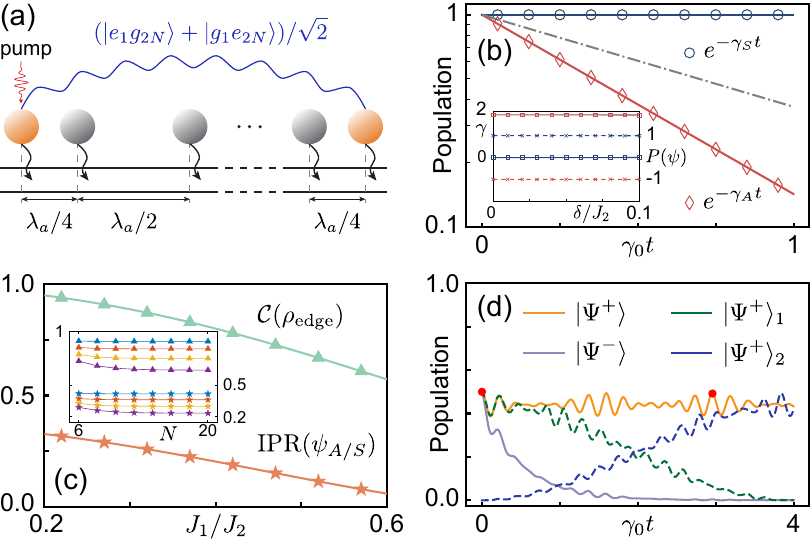}
\caption{(a) Schematic of a structured atomic chain coupled to a 1D waveguide. (b) Effective decay rates for hybridized edge states. The solid line is calculated by using the master equation. The inset shows their parity and decay rates in the presence of disorder. (c) Entanglement of the edge atoms from highly localized edge states. The inset shows their asymptotic limit with $N$ for $J_1/J_2=0.3$ (blue), $0.4$ (orange), $0.5$ (yellow), $0.6$ (purple), respectively. (d) Generation and transfer of the remote entanglement. The solid line shows the population of Bell states in a single SSH chain, where the red points correspond to concurrence $C_0=0$, $C_\mathrm{max}=0.49$. The dashed line shows the transfer of Bell state between two topological chains. Here, $J_1=0.25J_2$, $\gamma_0=0.035J_2$, $N=5$ and $N=3$ for the solid and dashed lines, respectively.}
\label{Fig4}
\end{figure}

\emph{Parity-associated super/subradiance and remote entanglement.}---In practice, the quantum system is inevitably affected by the environment, and this can be characterized by a non-unitary evolution. Here, we utilize a 1D waveguide to confine the radiative emission from atoms. The interference of the collective radiation can suppress the decay of hybridized edge states through the waveguide bath~\cite{bello2019unconventional,nie2020topologically,McDonnell2022subradiantedge}. 
As shown in Fig.~\ref{Fig1} (a), each atomic chain couples to an independent waveguide, which induces the dissipative couplings and gives rise to the correlated two-body dissipative dynamics~\cite{nie2021dissipative,lehmberg1970radiation,dung2002resonant,ficek2002entangled},
\begin{equation}
\dot{\rho}(t)=-\frac{i}{\hbar}\left[H_{0}+H_\mathrm{nl}, \rho(t)\right]+\mathcal{L}\rho,
\end{equation}
where $H_\mathrm{nl}=\sum_{l=1,2}\sum_{i, j} \hbar g_{i j}(\sigma_{l,i}^{+} \sigma_{l,j}^{-}+\sigma_{l,i}^{-} \sigma_{l,j}^{+})$ denotes the nonlocal interaction induced by the exchanging of virtual photons, and the Lindblad superoperator~\cite{le2005nanofiber,gonzalez2011entanglement,chang2012cavity,lalumiere2013input,doyeux2017giant,mirhosseini2019cavity,wen2019large}
\begin{equation}
\mathcal{L}\rho=\sum_{l=1,2}\sum_{i,j=1}^{2N}\gamma_{i j}\Big(\sigma_{li}^{-} \rho \sigma_{lj}^{+}-\frac{1}{2} \sigma_{li}^{+}\sigma_{lj}^{-} \rho  -\frac{1}{2} \rho \sigma_{li}^{+} \sigma_{lj}^{-}\Big)
\end{equation}
shows the correlated decay of the system. The non-local interaction and correlated decay are analytically given by~\cite{lu2023controllable} $g_{i j}=\gamma_{0} \sin \left(2 \pi d_{i j} / \lambda_{a}\right) / 2$ and $\gamma_{ij}=\gamma_0\cos{\left(2 \pi d_{i j} / \lambda_{a}\right)}$, respectively, where $\gamma_0$ is the spontaneous emission rate to the waveguide bath. 
The two-body interaction $g_{ij}$ and decay $\gamma_{ij}$ oscillate with the distance between atoms $d_{ij}$ scaled with the characteristic wavelength $\lambda_a=2\pi c/\omega_0$. We assume that the hopping rate $J_i$ is much larger than the atom-field coupling such that the interaction term plays a role of disorder. Here, we precisely design each atomic chain such that the distance between bulk atoms is $\lambda_a/2$, and the edge atoms are separated from bulk atoms by $\lambda_a/4$, as shown in Fig.~\ref{Fig4}(a).
In this case, the dissipative dynamics of edge atoms decouples from bulk atoms, accompanied by parity-associated super/subradiant states. The non-Hermitian Hamiltonian in the single-excitation subspace can be written as
\begin{equation}
H_\mathrm{nh} = H_\mathrm{SSH}  -i\Gamma_{E,-}|\Gamma_{E,-}\rangle\langle \Gamma_{E,-}| - i\Gamma_B|\Gamma_B\rangle\langle \Gamma_B|,
\end{equation}
where $|\Gamma_{E,-}\rangle$ and $|\Gamma_B\rangle$ denote the anti-symmetric dissipative eigenmodes for the edge and bulk atoms respectively, with decay rates $\Gamma_{E,-}=2\gamma_0$ and $\Gamma_B=(2N-2)\gamma_0$. The effective decay rates of the hybridized edge states are given by
\begin{equation}
\gamma_{A/S}=\Gamma_{E,-}|\langle\Gamma_{E,-}|\psi_{A/S}\rangle|^2 + \Gamma_{B} |\langle\Gamma_{B}|\psi_{A/S}\rangle|^2,
\end{equation}
with $\gamma_A=0$, $\gamma_S\simeq 2\gamma_0$. As shown in Fig.~\ref{Fig4}(b), the edge states with odd/even parity exhibits super/subradiance, which corresponds to a dissipative bright/dark state~\cite{van2013photon,asenjo2017atom,zhang2019theory,zanner2022coherent}. Moreover, in the inset of Fig.~\ref{Fig4}(b), we demonstrate the topological stability of the super/subradiance when considering the disorder, where the parity function is defined as $P(\psi)=\sum_{j=1}^{2N} \left|\langle\psi|\sigma_j^+|G\rangle + \langle\psi|I\sigma_j^+|G\rangle \right|^2-1$. 
Although the disorder breaks the inversion symmetry, the parity shows robustness since the edge states are protected by chiral symmetry of the SSH chain.

The similarity between highly localized edge states and Bell states $|\Psi^\pm\rangle=(|e_1 g_{2N}\rangle\pm |g_1 e_{2N}\rangle)/\sqrt 2$ also enables the generation of remote entangled states. By tracing out the bulk atoms, we investigate the concurrence $\mathcal{C}(\rho)$ of the reduced density matrix $\rho_\mathrm{edge}$ of the edge states. Fig.~\ref{Fig4}(c) shows the concurrence and the inverse participation ratio (IPR) of two edge atoms versus $J_1/J_2$, where the IPR measures the localization of wave functions, with $\mathrm{IPR}(\psi)=\sum_i\left|\psi\left(r_i\right)\right|^4=\sum_i p_i^2$ and $p_i$ being the probability at $i$-th site. Furthermore, they exhibit asymptotic independence as the cell number $N$ increases~\cite{lu2023controllable}. In order to prepare remote entanglement, one can first excite an edge atom via a pump field, as shown in Fig.~\ref{Fig4}(a). The entangled state $|\Psi^-\rangle$ decays quickly, while $|\Psi^+\rangle$ shows a slight oscillation with other dark states, around the steady value of mean concurrence $\bar{C}\simeq0.5|\langle\psi_S|\Psi^+\rangle|^4$. Fig.~\ref{Fig4}(d) displays the remote entanglement transfer in decoherence-free subspace between the two topological chains, through the tunable couplings of edge states, which also inherits their topological robustness.

\emph{Conclusion.}---In summary, we have proposed a topological system for studying controllable interaction via a local junction. The dynamics is constrained in the zero-energy manifold, providing two interacting subspaces for transitions of edge states. By tuning the junction interactions with $C_4$ and $\mathbb{Z}_2$ geometric symmetries, we find different interacting channels between edge states. According to the theory we implement the robust quantum state transfer and SWAP gate for the topological system, which suggests the excellent approximation from total Hamiltonian to the effective spin model. Beyond the coherent dynamics, this system shows the symmetric edge states are immune to environmental dissipation, which leads to protection against both disorder and decoherence. Our work paves the way towards controlling topological edge states with few-body local interactions in a non-adiabatic process.

	
\emph{Acknowledgments.}---We thank Y.-X. Liu, J.-Q. Liao and Z. Peng for stimulating discussions. This work is supported by the National Key R\&D Program of China (Grant No. 2019YFA0308200), the National Natural Science Foundation of China (Grant No. 11874432).
\bibliography{reference_XSSH.bib}

\begin{thebibliography}{65}%
\makeatletter
\providecommand \@ifxundefined [1]{%
 \@ifx{#1\undefined}
}%
\providecommand \@ifnum [1]{%
 \ifnum #1\expandafter \@firstoftwo
 \else \expandafter \@secondoftwo
 \fi
}%
\providecommand \@ifx [1]{%
 \ifx #1\expandafter \@firstoftwo
 \else \expandafter \@secondoftwo
 \fi
}%
\providecommand \natexlab [1]{#1}%
\providecommand \enquote  [1]{``#1''}%
\providecommand \bibnamefont  [1]{#1}%
\providecommand \bibfnamefont [1]{#1}%
\providecommand \citenamefont [1]{#1}%
\providecommand \href@noop [0]{\@secondoftwo}%
\providecommand \href [0]{\begingroup \@sanitize@url \@href}%
\providecommand \@href[1]{\@@startlink{#1}\@@href}%
\providecommand \@@href[1]{\endgroup#1\@@endlink}%
\providecommand \@sanitize@url [0]{\catcode `\\12\catcode `\$12\catcode
  `\&12\catcode `\#12\catcode `\^12\catcode `\_12\catcode `\%12\relax}%
\providecommand \@@startlink[1]{}%
\providecommand \@@endlink[0]{}%
\providecommand \url  [0]{\begingroup\@sanitize@url \@url }%
\providecommand \@url [1]{\endgroup\@href {#1}{\urlprefix }}%
\providecommand \urlprefix  [0]{URL }%
\providecommand \Eprint [0]{\href }%
\providecommand \doibase [0]{https://doi.org/}%
\providecommand \selectlanguage [0]{\@gobble}%
\providecommand \bibinfo  [0]{\@secondoftwo}%
\providecommand \bibfield  [0]{\@secondoftwo}%
\providecommand \translation [1]{[#1]}%
\providecommand \BibitemOpen [0]{}%
\providecommand \bibitemStop [0]{}%
\providecommand \bibitemNoStop [0]{.\EOS\space}%
\providecommand \EOS [0]{\spacefactor3000\relax}%
\providecommand \BibitemShut  [1]{\csname bibitem#1\endcsname}%
\let\auto@bib@innerbib\@empty
\bibitem [{\citenamefont {Hasan}\ and\ \citenamefont
  {Kane}(2010)}]{hasan2010colloquium}%
  \BibitemOpen
  \bibfield  {author} {\bibinfo {author} {\bibfnamefont {M.~Z.}\ \bibnamefont
  {Hasan}}\ and\ \bibinfo {author} {\bibfnamefont {C.~L.}\ \bibnamefont
  {Kane}},\ }\bibfield  {title} {\bibinfo {title} {Colloquium: Topological
  insulators},\ }\href {https://doi.org/10.1103/RevModPhys.82.3045} {\bibfield
  {journal} {\bibinfo  {journal} {Rev. Mod. Phys.}\ }\textbf {\bibinfo {volume}
  {82}},\ \bibinfo {pages} {3045} (\bibinfo {year} {2010})}\BibitemShut
  {NoStop}%
\bibitem [{\citenamefont {Qi}\ and\ \citenamefont
  {Zhang}(2011)}]{qi2011topological}%
  \BibitemOpen
  \bibfield  {author} {\bibinfo {author} {\bibfnamefont {X.-L.}\ \bibnamefont
  {Qi}}\ and\ \bibinfo {author} {\bibfnamefont {S.-C.}\ \bibnamefont {Zhang}},\
  }\bibfield  {title} {\bibinfo {title} {Topological insulators and
  superconductors},\ }\href {https://doi.org/10.1103/RevModPhys.83.1057}
  {\bibfield  {journal} {\bibinfo  {journal} {Rev. Mod. Phys.}\ }\textbf
  {\bibinfo {volume} {83}},\ \bibinfo {pages} {1057} (\bibinfo {year}
  {2011})}\BibitemShut {NoStop}%
\bibitem [{\citenamefont {Halperin}(1982)}]{halperin1982quantized}%
  \BibitemOpen
  \bibfield  {author} {\bibinfo {author} {\bibfnamefont {B.~I.}\ \bibnamefont
  {Halperin}},\ }\bibfield  {title} {\bibinfo {title} {Quantized {Hall}
  conductance, current-carrying edge states, and the existence of extended
  states in a two-dimensional disordered potential},\ }\href
  {https://doi.org/10.1103/PhysRevB.25.2185} {\bibfield  {journal} {\bibinfo
  {journal} {Phys. Rev. B}\ }\textbf {\bibinfo {volume} {25}},\ \bibinfo
  {pages} {2185} (\bibinfo {year} {1982})}\BibitemShut {NoStop}%
\bibitem [{\citenamefont {Niu}\ and\ \citenamefont
  {Thouless}(1984)}]{niu1984quantised}%
  \BibitemOpen
  \bibfield  {author} {\bibinfo {author} {\bibfnamefont {Q.}~\bibnamefont
  {Niu}}\ and\ \bibinfo {author} {\bibfnamefont {D.~J.}\ \bibnamefont
  {Thouless}},\ }\bibfield  {title} {\bibinfo {title} {Quantised adiabatic
  charge transport in the presence of substrate disorder and many-body
  interaction},\ }\href {https://doi.org/10.1088/0305-4470/17/12/016}
  {\bibfield  {journal} {\bibinfo  {journal} {Journal of Physics A:
  Mathematical and General}\ }\textbf {\bibinfo {volume} {17}},\ \bibinfo
  {pages} {2453} (\bibinfo {year} {1984})}\BibitemShut {NoStop}%
\bibitem [{\citenamefont {B\"uttiker}(1988)}]{buttiker1988absence}%
  \BibitemOpen
  \bibfield  {author} {\bibinfo {author} {\bibfnamefont {M.}~\bibnamefont
  {B\"uttiker}},\ }\bibfield  {title} {\bibinfo {title} {Absence of
  backscattering in the quantum {Hall} effect in multiprobe conductors},\
  }\href {https://doi.org/10.1103/PhysRevB.38.9375} {\bibfield  {journal}
  {\bibinfo  {journal} {Phys. Rev. B}\ }\textbf {\bibinfo {volume} {38}},\
  \bibinfo {pages} {9375} (\bibinfo {year} {1988})}\BibitemShut {NoStop}%
\bibitem [{\citenamefont {Hafezi}\ \emph {et~al.}(2011)\citenamefont {Hafezi},
  \citenamefont {Demler}, \citenamefont {Lukin},\ and\ \citenamefont
  {Taylor}}]{Hafezi2011}%
  \BibitemOpen
  \bibfield  {author} {\bibinfo {author} {\bibfnamefont {M.}~\bibnamefont
  {Hafezi}}, \bibinfo {author} {\bibfnamefont {E.~A.}\ \bibnamefont {Demler}},
  \bibinfo {author} {\bibfnamefont {M.~D.}\ \bibnamefont {Lukin}},\ and\
  \bibinfo {author} {\bibfnamefont {J.~M.}\ \bibnamefont {Taylor}},\ }\bibfield
   {title} {\bibinfo {title} {Robust optical delay lines with topological
  protection},\ }\href {https://doi.org/10.1038/nphys2063} {\bibfield
  {journal} {\bibinfo  {journal} {Nat. Phys.}\ }\textbf {\bibinfo {volume}
  {7}},\ \bibinfo {pages} {907} (\bibinfo {year} {2011})}\BibitemShut {NoStop}%
\bibitem [{\citenamefont {Perczel}\ \emph {et~al.}(2017)\citenamefont
  {Perczel}, \citenamefont {Borregaard}, \citenamefont {Chang}, \citenamefont
  {Pichler}, \citenamefont {Yelin}, \citenamefont {Zoller},\ and\ \citenamefont
  {Lukin}}]{perczel2017topological}%
  \BibitemOpen
  \bibfield  {author} {\bibinfo {author} {\bibfnamefont {J.}~\bibnamefont
  {Perczel}}, \bibinfo {author} {\bibfnamefont {J.}~\bibnamefont {Borregaard}},
  \bibinfo {author} {\bibfnamefont {D.~E.}\ \bibnamefont {Chang}}, \bibinfo
  {author} {\bibfnamefont {H.}~\bibnamefont {Pichler}}, \bibinfo {author}
  {\bibfnamefont {S.~F.}\ \bibnamefont {Yelin}}, \bibinfo {author}
  {\bibfnamefont {P.}~\bibnamefont {Zoller}},\ and\ \bibinfo {author}
  {\bibfnamefont {M.~D.}\ \bibnamefont {Lukin}},\ }\bibfield  {title} {\bibinfo
  {title} {Topological quantum optics in two-dimensional atomic arrays},\
  }\href {https://doi.org/10.1103/PhysRevLett.119.023603} {\bibfield  {journal}
  {\bibinfo  {journal} {Phys. Rev. Lett.}\ }\textbf {\bibinfo {volume} {119}},\
  \bibinfo {pages} {023603} (\bibinfo {year} {2017})}\BibitemShut {NoStop}%
\bibitem [{\citenamefont {Bandres}\ \emph {et~al.}(2018)\citenamefont
  {Bandres}, \citenamefont {Wittek}, \citenamefont {Harari}, \citenamefont
  {Parto}, \citenamefont {Ren}, \citenamefont {Segev}, \citenamefont
  {Christodoulides},\ and\ \citenamefont
  {Khajavikhan}}]{bandres2018topological}%
  \BibitemOpen
  \bibfield  {author} {\bibinfo {author} {\bibfnamefont {M.~A.}\ \bibnamefont
  {Bandres}}, \bibinfo {author} {\bibfnamefont {S.}~\bibnamefont {Wittek}},
  \bibinfo {author} {\bibfnamefont {G.}~\bibnamefont {Harari}}, \bibinfo
  {author} {\bibfnamefont {M.}~\bibnamefont {Parto}}, \bibinfo {author}
  {\bibfnamefont {J.}~\bibnamefont {Ren}}, \bibinfo {author} {\bibfnamefont
  {M.}~\bibnamefont {Segev}}, \bibinfo {author} {\bibfnamefont {D.~N.}\
  \bibnamefont {Christodoulides}},\ and\ \bibinfo {author} {\bibfnamefont
  {M.}~\bibnamefont {Khajavikhan}},\ }\bibfield  {title} {\bibinfo {title}
  {Topological insulator laser: Experiments},\ }\href
  {https://doi.org/10.1126/science.aar4005} {\bibfield  {journal} {\bibinfo
  {journal} {Science}\ }\textbf {\bibinfo {volume} {359}},\ \bibinfo {pages}
  {eaar4005} (\bibinfo {year} {2018})}\BibitemShut {NoStop}%
\bibitem [{\citenamefont {Ryu}\ and\ \citenamefont
  {Hatsugai}(2002)}]{ryu2002topological}%
  \BibitemOpen
  \bibfield  {author} {\bibinfo {author} {\bibfnamefont {S.}~\bibnamefont
  {Ryu}}\ and\ \bibinfo {author} {\bibfnamefont {Y.}~\bibnamefont {Hatsugai}},\
  }\bibfield  {title} {\bibinfo {title} {Topological origin of zero-energy edge
  states in particle-hole symmetric systems},\ }\href
  {https://doi.org/10.1103/PhysRevLett.89.077002} {\bibfield  {journal}
  {\bibinfo  {journal} {Phys. Rev. Lett.}\ }\textbf {\bibinfo {volume} {89}},\
  \bibinfo {pages} {077002} (\bibinfo {year} {2002})}\BibitemShut {NoStop}%
\bibitem [{\citenamefont {Asb{\'o}th}\ \emph {et~al.}(2016)\citenamefont
  {Asb{\'o}th}, \citenamefont {Oroszl{\'a}ny},\ and\ \citenamefont
  {P{\'a}lyi}}]{asboth2016short}%
  \BibitemOpen
  \bibfield  {author} {\bibinfo {author} {\bibfnamefont {J.~K.}\ \bibnamefont
  {Asb{\'o}th}}, \bibinfo {author} {\bibfnamefont {L.}~\bibnamefont
  {Oroszl{\'a}ny}},\ and\ \bibinfo {author} {\bibfnamefont {A.}~\bibnamefont
  {P{\'a}lyi}},\ }\href {https://doi.org/10.1007/978-3-319-25607-8} {\emph
  {\bibinfo {title} {A Short Course on Topological Insulators}}}\ (\bibinfo
  {publisher} {Springer International Publishing},\ \bibinfo {year}
  {2016})\BibitemShut {NoStop}%
\bibitem [{\citenamefont {Nie}\ and\ \citenamefont
  {Liu}(2020)}]{nie2020bandgap}%
  \BibitemOpen
  \bibfield  {author} {\bibinfo {author} {\bibfnamefont {W.}~\bibnamefont
  {Nie}}\ and\ \bibinfo {author} {\bibfnamefont {Y.-x.}\ \bibnamefont {Liu}},\
  }\bibfield  {title} {\bibinfo {title} {Bandgap-assisted quantum control of
  topological edge states in a cavity},\ }\href
  {https://doi.org/10.1103/PhysRevResearch.2.012076} {\bibfield  {journal}
  {\bibinfo  {journal} {Phys. Rev. Res.}\ }\textbf {\bibinfo {volume} {2}},\
  \bibinfo {pages} {012076(R)} (\bibinfo {year} {2020})}\BibitemShut {NoStop}%
\bibitem [{\citenamefont {Yao}\ \emph {et~al.}(2013)\citenamefont {Yao},
  \citenamefont {Laumann}, \citenamefont {Gorshkov}, \citenamefont {Weimer},
  \citenamefont {Jiang}, \citenamefont {Cirac}, \citenamefont {Zoller},\ and\
  \citenamefont {Lukin}}]{yao2013topologically}%
  \BibitemOpen
  \bibfield  {author} {\bibinfo {author} {\bibfnamefont {N.~Y.}\ \bibnamefont
  {Yao}}, \bibinfo {author} {\bibfnamefont {C.~R.}\ \bibnamefont {Laumann}},
  \bibinfo {author} {\bibfnamefont {A.~V.}\ \bibnamefont {Gorshkov}}, \bibinfo
  {author} {\bibfnamefont {H.}~\bibnamefont {Weimer}}, \bibinfo {author}
  {\bibfnamefont {L.}~\bibnamefont {Jiang}}, \bibinfo {author} {\bibfnamefont
  {J.~I.}\ \bibnamefont {Cirac}}, \bibinfo {author} {\bibfnamefont
  {P.}~\bibnamefont {Zoller}},\ and\ \bibinfo {author} {\bibfnamefont {M.~D.}\
  \bibnamefont {Lukin}},\ }\bibfield  {title} {\bibinfo {title} {Topologically
  protected quantum state transfer in a chiral spin liquid},\ }\href
  {https://doi.org/10.1038/ncomms2531} {\bibfield  {journal} {\bibinfo
  {journal} {Nature Communications}\ }\textbf {\bibinfo {volume} {4}},\
  \bibinfo {pages} {1585} (\bibinfo {year} {2013})}\BibitemShut {NoStop}%
\bibitem [{\citenamefont {Dlaska}\ \emph {et~al.}(2017)\citenamefont {Dlaska},
  \citenamefont {Vermersch},\ and\ \citenamefont {Zoller}}]{Dlaska_2017}%
  \BibitemOpen
  \bibfield  {author} {\bibinfo {author} {\bibfnamefont {C.}~\bibnamefont
  {Dlaska}}, \bibinfo {author} {\bibfnamefont {B.}~\bibnamefont {Vermersch}},\
  and\ \bibinfo {author} {\bibfnamefont {P.}~\bibnamefont {Zoller}},\
  }\bibfield  {title} {\bibinfo {title} {Robust quantum state transfer via
  topologically protected edge channels in dipolar arrays},\ }\href
  {https://doi.org/10.1088/2058-9565/2/1/015001} {\bibfield  {journal}
  {\bibinfo  {journal} {Quantum Science and Technology}\ }\textbf {\bibinfo
  {volume} {2}},\ \bibinfo {pages} {015001} (\bibinfo {year}
  {2017})}\BibitemShut {NoStop}%
\bibitem [{\citenamefont {Lang}\ and\ \citenamefont
  {B{\"u}chler}(2017)}]{lang2017topological}%
  \BibitemOpen
  \bibfield  {author} {\bibinfo {author} {\bibfnamefont {N.}~\bibnamefont
  {Lang}}\ and\ \bibinfo {author} {\bibfnamefont {H.~P.}\ \bibnamefont
  {B{\"u}chler}},\ }\bibfield  {title} {\bibinfo {title} {Topological networks
  for quantum communication between distant qubits},\ }\href
  {https://doi.org/10.1038/s41534-017-0047-x} {\bibfield  {journal} {\bibinfo
  {journal} {npj Quantum Information}\ }\textbf {\bibinfo {volume} {3}},\
  \bibinfo {pages} {47} (\bibinfo {year} {2017})}\BibitemShut {NoStop}%
\bibitem [{\citenamefont {Mei}\ \emph {et~al.}(2018)\citenamefont {Mei},
  \citenamefont {Chen}, \citenamefont {Tian}, \citenamefont {Zhu},\ and\
  \citenamefont {Jia}}]{mei2018robust}%
  \BibitemOpen
  \bibfield  {author} {\bibinfo {author} {\bibfnamefont {F.}~\bibnamefont
  {Mei}}, \bibinfo {author} {\bibfnamefont {G.}~\bibnamefont {Chen}}, \bibinfo
  {author} {\bibfnamefont {L.}~\bibnamefont {Tian}}, \bibinfo {author}
  {\bibfnamefont {S.-L.}\ \bibnamefont {Zhu}},\ and\ \bibinfo {author}
  {\bibfnamefont {S.}~\bibnamefont {Jia}},\ }\bibfield  {title} {\bibinfo
  {title} {Robust quantum state transfer via topological edge states in
  superconducting qubit chains},\ }\href
  {https://doi.org/10.1103/PhysRevA.98.012331} {\bibfield  {journal} {\bibinfo
  {journal} {Phys. Rev. A}\ }\textbf {\bibinfo {volume} {98}},\ \bibinfo
  {pages} {012331} (\bibinfo {year} {2018})}\BibitemShut {NoStop}%
\bibitem [{\citenamefont {Longhi}\ \emph {et~al.}(2019)\citenamefont {Longhi},
  \citenamefont {Giorgi},\ and\ \citenamefont {Zambrini}}]{longhi2019landau}%
  \BibitemOpen
  \bibfield  {author} {\bibinfo {author} {\bibfnamefont {S.}~\bibnamefont
  {Longhi}}, \bibinfo {author} {\bibfnamefont {G.~L.}\ \bibnamefont {Giorgi}},\
  and\ \bibinfo {author} {\bibfnamefont {R.}~\bibnamefont {Zambrini}},\
  }\bibfield  {title} {\bibinfo {title} {Landau–zener topological quantum
  state transfer},\ }\href {https://doi.org/10.1002/qute.201800090} {\bibfield
  {journal} {\bibinfo  {journal} {Advanced Quantum Technologies}\ }\textbf
  {\bibinfo {volume} {2}},\ \bibinfo {pages} {1800090} (\bibinfo {year}
  {2019})}\BibitemShut {NoStop}%
\bibitem [{\citenamefont {D'Angelis}\ \emph {et~al.}(2020)\citenamefont
  {D'Angelis}, \citenamefont {Pinheiro}, \citenamefont {Gu\'ery-Odelin},
  \citenamefont {Longhi},\ and\ \citenamefont {Impens}}]{d2020fast}%
  \BibitemOpen
  \bibfield  {author} {\bibinfo {author} {\bibfnamefont {F.~M.}\ \bibnamefont
  {D'Angelis}}, \bibinfo {author} {\bibfnamefont {F.~A.}\ \bibnamefont
  {Pinheiro}}, \bibinfo {author} {\bibfnamefont {D.}~\bibnamefont
  {Gu\'ery-Odelin}}, \bibinfo {author} {\bibfnamefont {S.}~\bibnamefont
  {Longhi}},\ and\ \bibinfo {author} {\bibfnamefont {F.}~\bibnamefont
  {Impens}},\ }\bibfield  {title} {\bibinfo {title} {Fast and robust quantum
  state transfer in a topological {Su}-{Schrieffer}-{Heeger} chain with
  next-to-nearest-neighbor interactions},\ }\href
  {https://doi.org/10.1103/PhysRevResearch.2.033475} {\bibfield  {journal}
  {\bibinfo  {journal} {Phys. Rev. Res.}\ }\textbf {\bibinfo {volume} {2}},\
  \bibinfo {pages} {033475} (\bibinfo {year} {2020})}\BibitemShut {NoStop}%
\bibitem [{\citenamefont {Qi}\ \emph {et~al.}(2021)\citenamefont {Qi},
  \citenamefont {Yan}, \citenamefont {Xing}, \citenamefont {Zhao},
  \citenamefont {Liu}, \citenamefont {Cui}, \citenamefont {Han}, \citenamefont
  {Zhang},\ and\ \citenamefont {Wang}}]{qi2021topological}%
  \BibitemOpen
  \bibfield  {author} {\bibinfo {author} {\bibfnamefont {L.}~\bibnamefont
  {Qi}}, \bibinfo {author} {\bibfnamefont {Y.}~\bibnamefont {Yan}}, \bibinfo
  {author} {\bibfnamefont {Y.}~\bibnamefont {Xing}}, \bibinfo {author}
  {\bibfnamefont {X.-D.}\ \bibnamefont {Zhao}}, \bibinfo {author}
  {\bibfnamefont {S.}~\bibnamefont {Liu}}, \bibinfo {author} {\bibfnamefont
  {W.-X.}\ \bibnamefont {Cui}}, \bibinfo {author} {\bibfnamefont
  {X.}~\bibnamefont {Han}}, \bibinfo {author} {\bibfnamefont {S.}~\bibnamefont
  {Zhang}},\ and\ \bibinfo {author} {\bibfnamefont {H.-F.}\ \bibnamefont
  {Wang}},\ }\bibfield  {title} {\bibinfo {title} {Topological router induced
  via long-range hopping in a {Su}-{Schrieffer}-{Heeger} chain},\ }\href
  {https://doi.org/10.1103/PhysRevResearch.3.023037} {\bibfield  {journal}
  {\bibinfo  {journal} {Phys. Rev. Research}\ }\textbf {\bibinfo {volume}
  {3}},\ \bibinfo {pages} {023037} (\bibinfo {year} {2021})}\BibitemShut
  {NoStop}%
\bibitem [{\citenamefont {Alicea}\ \emph {et~al.}(2011)\citenamefont {Alicea},
  \citenamefont {Oreg}, \citenamefont {Refael}, \citenamefont {von Oppen},\
  and\ \citenamefont {Fisher}}]{Alicea2011}%
  \BibitemOpen
  \bibfield  {author} {\bibinfo {author} {\bibfnamefont {J.}~\bibnamefont
  {Alicea}}, \bibinfo {author} {\bibfnamefont {Y.}~\bibnamefont {Oreg}},
  \bibinfo {author} {\bibfnamefont {G.}~\bibnamefont {Refael}}, \bibinfo
  {author} {\bibfnamefont {F.}~\bibnamefont {von Oppen}},\ and\ \bibinfo
  {author} {\bibfnamefont {M.~P.~A.}\ \bibnamefont {Fisher}},\ }\bibfield
  {title} {\bibinfo {title} {Non-{Abelian} statistics and topological quantum
  information processing in {1D} wire networks},\ }\href
  {https://doi.org/10.1038/nphys1915} {\bibfield  {journal} {\bibinfo
  {journal} {Nature Physics}\ }\textbf {\bibinfo {volume} {7}},\ \bibinfo
  {pages} {412} (\bibinfo {year} {2011})}\BibitemShut {NoStop}%
\bibitem [{\citenamefont {Sarma}\ \emph {et~al.}(2015)\citenamefont {Sarma},
  \citenamefont {Freedman},\ and\ \citenamefont {Nayak}}]{sarma2015majorana}%
  \BibitemOpen
  \bibfield  {author} {\bibinfo {author} {\bibfnamefont {S.~D.}\ \bibnamefont
  {Sarma}}, \bibinfo {author} {\bibfnamefont {M.}~\bibnamefont {Freedman}},\
  and\ \bibinfo {author} {\bibfnamefont {C.}~\bibnamefont {Nayak}},\ }\bibfield
   {title} {\bibinfo {title} {Majorana zero modes and topological quantum
  computation},\ }\href {https://doi.org/10.1038/npjqi.2015.1} {\bibfield
  {journal} {\bibinfo  {journal} {npj Quantum Information}\ }\textbf {\bibinfo
  {volume} {1}},\ \bibinfo {pages} {15001} (\bibinfo {year}
  {2015})}\BibitemShut {NoStop}%
\bibitem [{\citenamefont {Karzig}\ \emph {et~al.}(2017)\citenamefont {Karzig},
  \citenamefont {Knapp}, \citenamefont {Lutchyn}, \citenamefont {Bonderson},
  \citenamefont {Hastings}, \citenamefont {Nayak}, \citenamefont {Alicea},
  \citenamefont {Flensberg}, \citenamefont {Plugge}, \citenamefont {Oreg},
  \citenamefont {Marcus},\ and\ \citenamefont {Freedman}}]{karzig2017scalable}%
  \BibitemOpen
  \bibfield  {author} {\bibinfo {author} {\bibfnamefont {T.}~\bibnamefont
  {Karzig}}, \bibinfo {author} {\bibfnamefont {C.}~\bibnamefont {Knapp}},
  \bibinfo {author} {\bibfnamefont {R.~M.}\ \bibnamefont {Lutchyn}}, \bibinfo
  {author} {\bibfnamefont {P.}~\bibnamefont {Bonderson}}, \bibinfo {author}
  {\bibfnamefont {M.~B.}\ \bibnamefont {Hastings}}, \bibinfo {author}
  {\bibfnamefont {C.}~\bibnamefont {Nayak}}, \bibinfo {author} {\bibfnamefont
  {J.}~\bibnamefont {Alicea}}, \bibinfo {author} {\bibfnamefont
  {K.}~\bibnamefont {Flensberg}}, \bibinfo {author} {\bibfnamefont
  {S.}~\bibnamefont {Plugge}}, \bibinfo {author} {\bibfnamefont
  {Y.}~\bibnamefont {Oreg}}, \bibinfo {author} {\bibfnamefont {C.~M.}\
  \bibnamefont {Marcus}},\ and\ \bibinfo {author} {\bibfnamefont {M.~H.}\
  \bibnamefont {Freedman}},\ }\bibfield  {title} {\bibinfo {title} {Scalable
  designs for quasiparticle-poisoning-protected topological quantum computation
  with {Majorana} zero modes},\ }\href
  {https://doi.org/10.1103/PhysRevB.95.235305} {\bibfield  {journal} {\bibinfo
  {journal} {Phys. Rev. B}\ }\textbf {\bibinfo {volume} {95}},\ \bibinfo
  {pages} {235305} (\bibinfo {year} {2017})}\BibitemShut {NoStop}%
\bibitem [{\citenamefont {Brouzos}\ \emph {et~al.}(2020)\citenamefont
  {Brouzos}, \citenamefont {Kiorpelidis}, \citenamefont {Diakonos},\ and\
  \citenamefont {Theocharis}}]{brouzos2020fast}%
  \BibitemOpen
  \bibfield  {author} {\bibinfo {author} {\bibfnamefont {I.}~\bibnamefont
  {Brouzos}}, \bibinfo {author} {\bibfnamefont {I.}~\bibnamefont
  {Kiorpelidis}}, \bibinfo {author} {\bibfnamefont {F.~K.}\ \bibnamefont
  {Diakonos}},\ and\ \bibinfo {author} {\bibfnamefont {G.}~\bibnamefont
  {Theocharis}},\ }\bibfield  {title} {\bibinfo {title} {Fast, robust, and
  amplified transfer of topological edge modes on a time-varying mechanical
  chain},\ }\href {https://doi.org/10.1103/PhysRevB.102.174312} {\bibfield
  {journal} {\bibinfo  {journal} {Phys. Rev. B}\ }\textbf {\bibinfo {volume}
  {102}},\ \bibinfo {pages} {174312} (\bibinfo {year} {2020})}\BibitemShut
  {NoStop}%
\bibitem [{\citenamefont {Tian}\ \emph {et~al.}(2022)\citenamefont {Tian},
  \citenamefont {Zhang}, \citenamefont {Zhang}, \citenamefont {Wu},
  \citenamefont {Lin}, \citenamefont {Zhou}, \citenamefont {Duan},
  \citenamefont {Jiang},\ and\ \citenamefont {Du}}]{tian2022experimental}%
  \BibitemOpen
  \bibfield  {author} {\bibinfo {author} {\bibfnamefont {T.}~\bibnamefont
  {Tian}}, \bibinfo {author} {\bibfnamefont {Y.}~\bibnamefont {Zhang}},
  \bibinfo {author} {\bibfnamefont {L.}~\bibnamefont {Zhang}}, \bibinfo
  {author} {\bibfnamefont {L.}~\bibnamefont {Wu}}, \bibinfo {author}
  {\bibfnamefont {S.}~\bibnamefont {Lin}}, \bibinfo {author} {\bibfnamefont
  {J.}~\bibnamefont {Zhou}}, \bibinfo {author} {\bibfnamefont {C.-K.}\
  \bibnamefont {Duan}}, \bibinfo {author} {\bibfnamefont {J.-H.}\ \bibnamefont
  {Jiang}},\ and\ \bibinfo {author} {\bibfnamefont {J.}~\bibnamefont {Du}},\
  }\bibfield  {title} {\bibinfo {title} {Experimental realization of
  nonreciprocal adiabatic transfer of phonons in a dynamically modulated
  nanomechanical topological insulator},\ }\href
  {https://doi.org/10.1103/PhysRevLett.129.215901} {\bibfield  {journal}
  {\bibinfo  {journal} {Phys. Rev. Lett.}\ }\textbf {\bibinfo {volume} {129}},\
  \bibinfo {pages} {215901} (\bibinfo {year} {2022})}\BibitemShut {NoStop}%
\bibitem [{\citenamefont {Pakkiam}\ \emph {et~al.}(2023)\citenamefont
  {Pakkiam}, \citenamefont {Kumar}, \citenamefont {Pletyukhov},\ and\
  \citenamefont {Fedorov}}]{Pakkiam2023}%
  \BibitemOpen
  \bibfield  {author} {\bibinfo {author} {\bibfnamefont {P.}~\bibnamefont
  {Pakkiam}}, \bibinfo {author} {\bibfnamefont {N.~P.}\ \bibnamefont {Kumar}},
  \bibinfo {author} {\bibfnamefont {M.}~\bibnamefont {Pletyukhov}},\ and\
  \bibinfo {author} {\bibfnamefont {A.}~\bibnamefont {Fedorov}},\ }\bibfield
  {title} {\bibinfo {title} {Qubit-controlled directional edge states in
  waveguide {QED}},\ }\href {https://doi.org/10.1038/s41534-023-00722-8}
  {\bibfield  {journal} {\bibinfo  {journal} {npj Quantum Information}\
  }\textbf {\bibinfo {volume} {9}},\ \bibinfo {pages} {53} (\bibinfo {year}
  {2023})}\BibitemShut {NoStop}%
\bibitem [{\citenamefont {Thouless}(1983)}]{thouless1983quantization}%
  \BibitemOpen
  \bibfield  {author} {\bibinfo {author} {\bibfnamefont {D.~J.}\ \bibnamefont
  {Thouless}},\ }\bibfield  {title} {\bibinfo {title} {Quantization of particle
  transport},\ }\href {https://doi.org/10.1103/PhysRevB.27.6083} {\bibfield
  {journal} {\bibinfo  {journal} {Phys. Rev. B}\ }\textbf {\bibinfo {volume}
  {27}},\ \bibinfo {pages} {6083} (\bibinfo {year} {1983})}\BibitemShut
  {NoStop}%
\bibitem [{\citenamefont {Lohse}\ \emph {et~al.}(2016)\citenamefont {Lohse},
  \citenamefont {Schweizer}, \citenamefont {Zilberberg}, \citenamefont
  {Aidelsburger},\ and\ \citenamefont {Bloch}}]{lohse2016thouless}%
  \BibitemOpen
  \bibfield  {author} {\bibinfo {author} {\bibfnamefont {M.}~\bibnamefont
  {Lohse}}, \bibinfo {author} {\bibfnamefont {C.}~\bibnamefont {Schweizer}},
  \bibinfo {author} {\bibfnamefont {O.}~\bibnamefont {Zilberberg}}, \bibinfo
  {author} {\bibfnamefont {M.}~\bibnamefont {Aidelsburger}},\ and\ \bibinfo
  {author} {\bibfnamefont {I.}~\bibnamefont {Bloch}},\ }\bibfield  {title}
  {\bibinfo {title} {A {Thouless} quantum pump with ultracold bosonic atoms in
  an optical superlattice},\ }\href {https://doi.org/10.1038/nphys3584}
  {\bibfield  {journal} {\bibinfo  {journal} {Nat. Phys.}\ }\textbf {\bibinfo
  {volume} {12}},\ \bibinfo {pages} {350} (\bibinfo {year} {2016})}\BibitemShut
  {NoStop}%
\bibitem [{\citenamefont {Nakajima}\ \emph {et~al.}(2016)\citenamefont
  {Nakajima}, \citenamefont {Tomita}, \citenamefont {Taie}, \citenamefont
  {Ichinose}, \citenamefont {Ozawa}, \citenamefont {Wang}, \citenamefont
  {Troyer},\ and\ \citenamefont {Takahashi}}]{nakajima2016topological}%
  \BibitemOpen
  \bibfield  {author} {\bibinfo {author} {\bibfnamefont {S.}~\bibnamefont
  {Nakajima}}, \bibinfo {author} {\bibfnamefont {T.}~\bibnamefont {Tomita}},
  \bibinfo {author} {\bibfnamefont {S.}~\bibnamefont {Taie}}, \bibinfo {author}
  {\bibfnamefont {T.}~\bibnamefont {Ichinose}}, \bibinfo {author}
  {\bibfnamefont {H.}~\bibnamefont {Ozawa}}, \bibinfo {author} {\bibfnamefont
  {L.}~\bibnamefont {Wang}}, \bibinfo {author} {\bibfnamefont {M.}~\bibnamefont
  {Troyer}},\ and\ \bibinfo {author} {\bibfnamefont {Y.}~\bibnamefont
  {Takahashi}},\ }\bibfield  {title} {\bibinfo {title} {Topological {Thouless}
  pumping of ultracold fermions},\ }\href {https://doi.org/10.1038/nphys3622}
  {\bibfield  {journal} {\bibinfo  {journal} {Nat. Phys.}\ }\textbf {\bibinfo
  {volume} {12}},\ \bibinfo {pages} {296} (\bibinfo {year} {2016})}\BibitemShut
  {NoStop}%
\bibitem [{\citenamefont {Gu}\ \emph {et~al.}(2017)\citenamefont {Gu},
  \citenamefont {Chen},\ and\ \citenamefont {xi~Liu}}]{gu2017topological}%
  \BibitemOpen
  \bibfield  {author} {\bibinfo {author} {\bibfnamefont {X.}~\bibnamefont
  {Gu}}, \bibinfo {author} {\bibfnamefont {S.}~\bibnamefont {Chen}},\ and\
  \bibinfo {author} {\bibfnamefont {Y.}~\bibnamefont {xi~Liu}},\ }\href@noop {}
  {\bibinfo {title} {Topological edge states and pumping in a chain of coupled
  superconducting qubits}} (\bibinfo {year} {2017}),\ \Eprint
  {https://arxiv.org/abs/1711.06829} {arXiv:1711.06829 [quant-ph]} \BibitemShut
  {NoStop}%
\bibitem [{\citenamefont {Longhi}(2019)}]{longhi2019topological}%
  \BibitemOpen
  \bibfield  {author} {\bibinfo {author} {\bibfnamefont {S.}~\bibnamefont
  {Longhi}},\ }\bibfield  {title} {\bibinfo {title} {Topological pumping of
  edge states via adiabatic passage},\ }\href
  {https://doi.org/10.1103/PhysRevB.99.155150} {\bibfield  {journal} {\bibinfo
  {journal} {Phys. Rev. B}\ }\textbf {\bibinfo {volume} {99}},\ \bibinfo
  {pages} {155150} (\bibinfo {year} {2019})}\BibitemShut {NoStop}%
\bibitem [{\citenamefont {Kraus}\ \emph {et~al.}(2012)\citenamefont {Kraus},
  \citenamefont {Lahini}, \citenamefont {Ringel}, \citenamefont {Verbin},\ and\
  \citenamefont {Zilberberg}}]{kraus2012topological}%
  \BibitemOpen
  \bibfield  {author} {\bibinfo {author} {\bibfnamefont {Y.~E.}\ \bibnamefont
  {Kraus}}, \bibinfo {author} {\bibfnamefont {Y.}~\bibnamefont {Lahini}},
  \bibinfo {author} {\bibfnamefont {Z.}~\bibnamefont {Ringel}}, \bibinfo
  {author} {\bibfnamefont {M.}~\bibnamefont {Verbin}},\ and\ \bibinfo {author}
  {\bibfnamefont {O.}~\bibnamefont {Zilberberg}},\ }\bibfield  {title}
  {\bibinfo {title} {Topological states and adiabatic pumping in
  quasicrystals},\ }\href {https://doi.org/10.1103/PhysRevLett.109.106402}
  {\bibfield  {journal} {\bibinfo  {journal} {Phys. Rev. Lett.}\ }\textbf
  {\bibinfo {volume} {109}},\ \bibinfo {pages} {106402} (\bibinfo {year}
  {2012})}\BibitemShut {NoStop}%
\bibitem [{\citenamefont {Verbin}\ \emph {et~al.}(2015)\citenamefont {Verbin},
  \citenamefont {Zilberberg}, \citenamefont {Lahini}, \citenamefont {Kraus},\
  and\ \citenamefont {Silberberg}}]{verbin2015topological}%
  \BibitemOpen
  \bibfield  {author} {\bibinfo {author} {\bibfnamefont {M.}~\bibnamefont
  {Verbin}}, \bibinfo {author} {\bibfnamefont {O.}~\bibnamefont {Zilberberg}},
  \bibinfo {author} {\bibfnamefont {Y.}~\bibnamefont {Lahini}}, \bibinfo
  {author} {\bibfnamefont {Y.~E.}\ \bibnamefont {Kraus}},\ and\ \bibinfo
  {author} {\bibfnamefont {Y.}~\bibnamefont {Silberberg}},\ }\bibfield  {title}
  {\bibinfo {title} {Topological pumping over a photonic {Fibonacci}
  quasicrystal},\ }\href {https://doi.org/10.1103/PhysRevB.91.064201}
  {\bibfield  {journal} {\bibinfo  {journal} {Phys. Rev. B}\ }\textbf {\bibinfo
  {volume} {91}},\ \bibinfo {pages} {064201} (\bibinfo {year}
  {2015})}\BibitemShut {NoStop}%
\bibitem [{\citenamefont {Zilberberg}\ \emph {et~al.}(2018)\citenamefont
  {Zilberberg}, \citenamefont {Huang}, \citenamefont {Guglielmon},
  \citenamefont {Wang}, \citenamefont {Chen}, \citenamefont {Kraus},\ and\
  \citenamefont {Rechtsman}}]{Zilberberg2018}%
  \BibitemOpen
  \bibfield  {author} {\bibinfo {author} {\bibfnamefont {O.}~\bibnamefont
  {Zilberberg}}, \bibinfo {author} {\bibfnamefont {S.}~\bibnamefont {Huang}},
  \bibinfo {author} {\bibfnamefont {J.}~\bibnamefont {Guglielmon}}, \bibinfo
  {author} {\bibfnamefont {M.}~\bibnamefont {Wang}}, \bibinfo {author}
  {\bibfnamefont {K.~P.}\ \bibnamefont {Chen}}, \bibinfo {author}
  {\bibfnamefont {Y.~E.}\ \bibnamefont {Kraus}},\ and\ \bibinfo {author}
  {\bibfnamefont {M.~C.}\ \bibnamefont {Rechtsman}},\ }\bibfield  {title}
  {\bibinfo {title} {Photonic topological boundary pumping as a probe of {4D}
  quantum {Hall} physics},\ }\href {https://doi.org/10.1038/nature25011}
  {\bibfield  {journal} {\bibinfo  {journal} {Nature}\ }\textbf {\bibinfo
  {volume} {553}},\ \bibinfo {pages} {59} (\bibinfo {year} {2018})}\BibitemShut
  {NoStop}%
\bibitem [{\citenamefont {Cheng}\ \emph {et~al.}(2022)\citenamefont {Cheng},
  \citenamefont {Wang}, \citenamefont {Ke}, \citenamefont {Chen}, \citenamefont
  {Yu}, \citenamefont {Kivshar}, \citenamefont {Lee},\ and\ \citenamefont
  {Pan}}]{Cheng2022}%
  \BibitemOpen
  \bibfield  {author} {\bibinfo {author} {\bibfnamefont {Q.}~\bibnamefont
  {Cheng}}, \bibinfo {author} {\bibfnamefont {H.}~\bibnamefont {Wang}},
  \bibinfo {author} {\bibfnamefont {Y.}~\bibnamefont {Ke}}, \bibinfo {author}
  {\bibfnamefont {T.}~\bibnamefont {Chen}}, \bibinfo {author} {\bibfnamefont
  {Y.}~\bibnamefont {Yu}}, \bibinfo {author} {\bibfnamefont {Y.~S.}\
  \bibnamefont {Kivshar}}, \bibinfo {author} {\bibfnamefont {C.}~\bibnamefont
  {Lee}},\ and\ \bibinfo {author} {\bibfnamefont {Y.}~\bibnamefont {Pan}},\
  }\bibfield  {title} {\bibinfo {title} {Asymmetric topological pumping in
  nonparaxial photonics},\ }\href {https://doi.org/10.1038/s41467-021-27773-9}
  {\bibfield  {journal} {\bibinfo  {journal} {Nature Communications}\ }\textbf
  {\bibinfo {volume} {13}},\ \bibinfo {pages} {249} (\bibinfo {year}
  {2022})}\BibitemShut {NoStop}%
\bibitem [{\citenamefont {Rosa}\ \emph {et~al.}(2019)\citenamefont {Rosa},
  \citenamefont {Pal}, \citenamefont {Arruda},\ and\ \citenamefont
  {Ruzzene}}]{rosa2019edge}%
  \BibitemOpen
  \bibfield  {author} {\bibinfo {author} {\bibfnamefont {M.~I.~N.}\
  \bibnamefont {Rosa}}, \bibinfo {author} {\bibfnamefont {R.~K.}\ \bibnamefont
  {Pal}}, \bibinfo {author} {\bibfnamefont {J.~R.~F.}\ \bibnamefont {Arruda}},\
  and\ \bibinfo {author} {\bibfnamefont {M.}~\bibnamefont {Ruzzene}},\
  }\bibfield  {title} {\bibinfo {title} {Edge states and topological pumping in
  spatially modulated elastic lattices},\ }\href
  {https://doi.org/10.1103/PhysRevLett.123.034301} {\bibfield  {journal}
  {\bibinfo  {journal} {Phys. Rev. Lett.}\ }\textbf {\bibinfo {volume} {123}},\
  \bibinfo {pages} {034301} (\bibinfo {year} {2019})}\BibitemShut {NoStop}%
\bibitem [{\citenamefont {Grinberg}\ \emph {et~al.}(2020)\citenamefont
  {Grinberg}, \citenamefont {Lin}, \citenamefont {Harris}, \citenamefont
  {Benalcazar}, \citenamefont {Peterson}, \citenamefont {Hughes},\ and\
  \citenamefont {Bahl}}]{Grinberg2020}%
  \BibitemOpen
  \bibfield  {author} {\bibinfo {author} {\bibfnamefont {I.~H.}\ \bibnamefont
  {Grinberg}}, \bibinfo {author} {\bibfnamefont {M.}~\bibnamefont {Lin}},
  \bibinfo {author} {\bibfnamefont {C.}~\bibnamefont {Harris}}, \bibinfo
  {author} {\bibfnamefont {W.~A.}\ \bibnamefont {Benalcazar}}, \bibinfo
  {author} {\bibfnamefont {C.~W.}\ \bibnamefont {Peterson}}, \bibinfo {author}
  {\bibfnamefont {T.~L.}\ \bibnamefont {Hughes}},\ and\ \bibinfo {author}
  {\bibfnamefont {G.}~\bibnamefont {Bahl}},\ }\bibfield  {title} {\bibinfo
  {title} {Robust temporal pumping in a magneto-mechanical topological
  insulator},\ }\href {https://doi.org/10.1038/s41467-020-14804-0} {\bibfield
  {journal} {\bibinfo  {journal} {Nature Communications}\ }\textbf {\bibinfo
  {volume} {11}},\ \bibinfo {pages} {974} (\bibinfo {year} {2020})}\BibitemShut
  {NoStop}%
\bibitem [{\citenamefont {Xia}\ \emph {et~al.}(2021)\citenamefont {Xia},
  \citenamefont {Riva}, \citenamefont {Rosa}, \citenamefont {Cazzulani},
  \citenamefont {Erturk}, \citenamefont {Braghin},\ and\ \citenamefont
  {Ruzzene}}]{xia2021experimental}%
  \BibitemOpen
  \bibfield  {author} {\bibinfo {author} {\bibfnamefont {Y.}~\bibnamefont
  {Xia}}, \bibinfo {author} {\bibfnamefont {E.}~\bibnamefont {Riva}}, \bibinfo
  {author} {\bibfnamefont {M.~I.~N.}\ \bibnamefont {Rosa}}, \bibinfo {author}
  {\bibfnamefont {G.}~\bibnamefont {Cazzulani}}, \bibinfo {author}
  {\bibfnamefont {A.}~\bibnamefont {Erturk}}, \bibinfo {author} {\bibfnamefont
  {F.}~\bibnamefont {Braghin}},\ and\ \bibinfo {author} {\bibfnamefont
  {M.}~\bibnamefont {Ruzzene}},\ }\bibfield  {title} {\bibinfo {title}
  {Experimental observation of temporal pumping in electromechanical
  waveguides},\ }\href {https://doi.org/10.1103/PhysRevLett.126.095501}
  {\bibfield  {journal} {\bibinfo  {journal} {Phys. Rev. Lett.}\ }\textbf
  {\bibinfo {volume} {126}},\ \bibinfo {pages} {095501} (\bibinfo {year}
  {2021})}\BibitemShut {NoStop}%
\bibitem [{\citenamefont {Cheng}\ \emph {et~al.}(2020)\citenamefont {Cheng},
  \citenamefont {Prodan},\ and\ \citenamefont
  {Prodan}}]{cheng2020experimental}%
  \BibitemOpen
  \bibfield  {author} {\bibinfo {author} {\bibfnamefont {W.}~\bibnamefont
  {Cheng}}, \bibinfo {author} {\bibfnamefont {E.}~\bibnamefont {Prodan}},\ and\
  \bibinfo {author} {\bibfnamefont {C.}~\bibnamefont {Prodan}},\ }\bibfield
  {title} {\bibinfo {title} {Experimental demonstration of dynamic topological
  pumping across incommensurate bilayered acoustic metamaterials},\ }\href
  {https://doi.org/10.1103/PhysRevLett.125.224301} {\bibfield  {journal}
  {\bibinfo  {journal} {Phys. Rev. Lett.}\ }\textbf {\bibinfo {volume} {125}},\
  \bibinfo {pages} {224301} (\bibinfo {year} {2020})}\BibitemShut {NoStop}%
\bibitem [{\citenamefont {Chen}\ \emph
  {et~al.}(2021{\natexlab{a}})\citenamefont {Chen}, \citenamefont {Tang},
  \citenamefont {Zhang}, \citenamefont {Chen},\ and\ \citenamefont
  {Ma}}]{chen2021landau}%
  \BibitemOpen
  \bibfield  {author} {\bibinfo {author} {\bibfnamefont {Z.-G.}\ \bibnamefont
  {Chen}}, \bibinfo {author} {\bibfnamefont {W.}~\bibnamefont {Tang}}, \bibinfo
  {author} {\bibfnamefont {R.-Y.}\ \bibnamefont {Zhang}}, \bibinfo {author}
  {\bibfnamefont {Z.}~\bibnamefont {Chen}},\ and\ \bibinfo {author}
  {\bibfnamefont {G.}~\bibnamefont {Ma}},\ }\bibfield  {title} {\bibinfo
  {title} {Landau-{Zener} transition in the dynamic transfer of acoustic
  topological states},\ }\href {https://doi.org/10.1103/PhysRevLett.126.054301}
  {\bibfield  {journal} {\bibinfo  {journal} {Phys. Rev. Lett.}\ }\textbf
  {\bibinfo {volume} {126}},\ \bibinfo {pages} {054301} (\bibinfo {year}
  {2021}{\natexlab{a}})}\BibitemShut {NoStop}%
\bibitem [{\citenamefont {Chen}\ \emph
  {et~al.}(2021{\natexlab{b}})\citenamefont {Chen}, \citenamefont {Zhang},
  \citenamefont {Wu}, \citenamefont {Huang}, \citenamefont {Nguyen},
  \citenamefont {Prodan}, \citenamefont {Zhou},\ and\ \citenamefont
  {Huang}}]{Chen2021}%
  \BibitemOpen
  \bibfield  {author} {\bibinfo {author} {\bibfnamefont {H.}~\bibnamefont
  {Chen}}, \bibinfo {author} {\bibfnamefont {H.}~\bibnamefont {Zhang}},
  \bibinfo {author} {\bibfnamefont {Q.}~\bibnamefont {Wu}}, \bibinfo {author}
  {\bibfnamefont {Y.}~\bibnamefont {Huang}}, \bibinfo {author} {\bibfnamefont
  {H.}~\bibnamefont {Nguyen}}, \bibinfo {author} {\bibfnamefont
  {E.}~\bibnamefont {Prodan}}, \bibinfo {author} {\bibfnamefont
  {X.}~\bibnamefont {Zhou}},\ and\ \bibinfo {author} {\bibfnamefont
  {G.}~\bibnamefont {Huang}},\ }\bibfield  {title} {\bibinfo {title} {Creating
  synthetic spaces for higher-order topological sound transport},\ }\href
  {https://doi.org/10.1038/s41467-021-25305-z} {\bibfield  {journal} {\bibinfo
  {journal} {Nature Communications}\ }\textbf {\bibinfo {volume} {12}},\
  \bibinfo {pages} {5028} (\bibinfo {year} {2021}{\natexlab{b}})}\BibitemShut
  {NoStop}%
\bibitem [{\citenamefont {Lahtinen}\ and\ \citenamefont
  {Pachos}(2017)}]{lahtinen2017short}%
  \BibitemOpen
  \bibfield  {author} {\bibinfo {author} {\bibfnamefont {V.}~\bibnamefont
  {Lahtinen}}\ and\ \bibinfo {author} {\bibfnamefont {J.~K.}\ \bibnamefont
  {Pachos}},\ }\bibfield  {title} {\bibinfo {title} {{A Short Introduction to
  Topological Quantum Computation}},\ }\href
  {https://doi.org/10.21468/SciPostPhys.3.3.021} {\bibfield  {journal}
  {\bibinfo  {journal} {SciPost Phys.}\ }\textbf {\bibinfo {volume} {3}},\
  \bibinfo {pages} {021} (\bibinfo {year} {2017})}\BibitemShut {NoStop}%
\bibitem [{\citenamefont {Boross}\ \emph {et~al.}(2019)\citenamefont {Boross},
  \citenamefont {Asb\'oth}, \citenamefont {Sz\'echenyi}, \citenamefont
  {Oroszl\'any},\ and\ \citenamefont {P\'alyi}}]{boross2019poor}%
  \BibitemOpen
  \bibfield  {author} {\bibinfo {author} {\bibfnamefont {P.}~\bibnamefont
  {Boross}}, \bibinfo {author} {\bibfnamefont {J.~K.}\ \bibnamefont
  {Asb\'oth}}, \bibinfo {author} {\bibfnamefont {G.}~\bibnamefont
  {Sz\'echenyi}}, \bibinfo {author} {\bibfnamefont {L.}~\bibnamefont
  {Oroszl\'any}},\ and\ \bibinfo {author} {\bibfnamefont {A.}~\bibnamefont
  {P\'alyi}},\ }\bibfield  {title} {\bibinfo {title} {Poor man's topological
  quantum gate based on the {Su}-{Schrieffer}-{Heeger} model},\ }\href
  {https://doi.org/10.1103/PhysRevB.100.045414} {\bibfield  {journal} {\bibinfo
   {journal} {Phys. Rev. B}\ }\textbf {\bibinfo {volume} {100}},\ \bibinfo
  {pages} {045414} (\bibinfo {year} {2019})}\BibitemShut {NoStop}%
\bibitem [{\citenamefont {Delplace}\ \emph {et~al.}(2011)\citenamefont
  {Delplace}, \citenamefont {Ullmo},\ and\ \citenamefont
  {Montambaux}}]{delplace2011zak}%
  \BibitemOpen
  \bibfield  {author} {\bibinfo {author} {\bibfnamefont {P.}~\bibnamefont
  {Delplace}}, \bibinfo {author} {\bibfnamefont {D.}~\bibnamefont {Ullmo}},\
  and\ \bibinfo {author} {\bibfnamefont {G.}~\bibnamefont {Montambaux}},\
  }\bibfield  {title} {\bibinfo {title} {Zak phase and the existence of edge
  states in graphene},\ }\href {https://doi.org/10.1103/PhysRevB.84.195452}
  {\bibfield  {journal} {\bibinfo  {journal} {Phys. Rev. B}\ }\textbf {\bibinfo
  {volume} {84}},\ \bibinfo {pages} {195452} (\bibinfo {year}
  {2011})}\BibitemShut {NoStop}%
\bibitem [{\citenamefont {Su}\ \emph {et~al.}(1979)\citenamefont {Su},
  \citenamefont {Schrieffer},\ and\ \citenamefont {Heeger}}]{su1979solitons}%
  \BibitemOpen
  \bibfield  {author} {\bibinfo {author} {\bibfnamefont {W.~P.}\ \bibnamefont
  {Su}}, \bibinfo {author} {\bibfnamefont {J.~R.}\ \bibnamefont {Schrieffer}},\
  and\ \bibinfo {author} {\bibfnamefont {A.~J.}\ \bibnamefont {Heeger}},\
  }\bibfield  {title} {\bibinfo {title} {Solitons in polyacetylene},\ }\href
  {https://doi.org/10.1103/PhysRevLett.42.1698} {\bibfield  {journal} {\bibinfo
   {journal} {Phys. Rev. Lett.}\ }\textbf {\bibinfo {volume} {42}},\ \bibinfo
  {pages} {1698} (\bibinfo {year} {1979})}\BibitemShut {NoStop}%
\bibitem [{\citenamefont {Xiang}\ \emph {et~al.}(2013)\citenamefont {Xiang},
  \citenamefont {Ashhab}, \citenamefont {You},\ and\ \citenamefont
  {Nori}}]{xiang2013hybrid}%
  \BibitemOpen
  \bibfield  {author} {\bibinfo {author} {\bibfnamefont {Z.-L.}\ \bibnamefont
  {Xiang}}, \bibinfo {author} {\bibfnamefont {S.}~\bibnamefont {Ashhab}},
  \bibinfo {author} {\bibfnamefont {J.~Q.}\ \bibnamefont {You}},\ and\ \bibinfo
  {author} {\bibfnamefont {F.}~\bibnamefont {Nori}},\ }\bibfield  {title}
  {\bibinfo {title} {Hybrid quantum circuits: Superconducting circuits
  interacting with other quantum systems},\ }\href
  {https://doi.org/10.1103/RevModPhys.85.623} {\bibfield  {journal} {\bibinfo
  {journal} {Rev. Mod. Phys.}\ }\textbf {\bibinfo {volume} {85}},\ \bibinfo
  {pages} {623} (\bibinfo {year} {2013})}\BibitemShut {NoStop}%
\bibitem [{lu2()}]{lu2023controllable}%
  \BibitemOpen
  \href@noop {} {\bibinfo {title} {See supplemental material for
  details}}\BibitemShut {NoStop}%
\bibitem [{\citenamefont {Kandel}\ \emph {et~al.}(2019)\citenamefont {Kandel},
  \citenamefont {Qiao}, \citenamefont {Fallahi}, \citenamefont {Gardner},
  \citenamefont {Manfra},\ and\ \citenamefont {Nichol}}]{kandel2019coherent}%
  \BibitemOpen
  \bibfield  {author} {\bibinfo {author} {\bibfnamefont {Y.~P.}\ \bibnamefont
  {Kandel}}, \bibinfo {author} {\bibfnamefont {H.}~\bibnamefont {Qiao}},
  \bibinfo {author} {\bibfnamefont {S.}~\bibnamefont {Fallahi}}, \bibinfo
  {author} {\bibfnamefont {G.~C.}\ \bibnamefont {Gardner}}, \bibinfo {author}
  {\bibfnamefont {M.~J.}\ \bibnamefont {Manfra}},\ and\ \bibinfo {author}
  {\bibfnamefont {J.~M.}\ \bibnamefont {Nichol}},\ }\bibfield  {title}
  {\bibinfo {title} {Coherent spin-state transfer via {Heisenberg} exchange},\
  }\href {https://doi.org/10.1038/s41586-019-1566-8} {\bibfield  {journal}
  {\bibinfo  {journal} {Nature}\ }\textbf {\bibinfo {volume} {573}},\ \bibinfo
  {pages} {553} (\bibinfo {year} {2019})}\BibitemShut {NoStop}%
\bibitem [{\citenamefont {Nielsen}(2002)}]{NIELSEN2002249}%
  \BibitemOpen
  \bibfield  {author} {\bibinfo {author} {\bibfnamefont {M.~A.}\ \bibnamefont
  {Nielsen}},\ }\bibfield  {title} {\bibinfo {title} {A simple formula for the
  average gate fidelity of a quantum dynamical operation},\ }\href
  {https://doi.org/https://doi.org/10.1016/S0375-9601(02)01272-0} {\bibfield
  {journal} {\bibinfo  {journal} {Physics Letters A}\ }\textbf {\bibinfo
  {volume} {303}},\ \bibinfo {pages} {249} (\bibinfo {year}
  {2002})}\BibitemShut {NoStop}%
\bibitem [{\citenamefont {Bello}\ \emph {et~al.}(2019)\citenamefont {Bello},
  \citenamefont {Platero}, \citenamefont {Cirac},\ and\ \citenamefont
  {González-Tudela}}]{bello2019unconventional}%
  \BibitemOpen
  \bibfield  {author} {\bibinfo {author} {\bibfnamefont {M.}~\bibnamefont
  {Bello}}, \bibinfo {author} {\bibfnamefont {G.}~\bibnamefont {Platero}},
  \bibinfo {author} {\bibfnamefont {J.~I.}\ \bibnamefont {Cirac}},\ and\
  \bibinfo {author} {\bibfnamefont {A.}~\bibnamefont {González-Tudela}},\
  }\bibfield  {title} {\bibinfo {title} {Unconventional quantum optics in
  topological waveguide {QED}},\ }\href
  {https://doi.org/10.1126/sciadv.aaw0297} {\bibfield  {journal} {\bibinfo
  {journal} {Science Advances}\ }\textbf {\bibinfo {volume} {5}},\ \bibinfo
  {pages} {eaaw0297} (\bibinfo {year} {2019})}\BibitemShut {NoStop}%
\bibitem [{\citenamefont {Nie}\ \emph {et~al.}(2020)\citenamefont {Nie},
  \citenamefont {Peng}, \citenamefont {Nori},\ and\ \citenamefont
  {Liu}}]{nie2020topologically}%
  \BibitemOpen
  \bibfield  {author} {\bibinfo {author} {\bibfnamefont {W.}~\bibnamefont
  {Nie}}, \bibinfo {author} {\bibfnamefont {Z.~H.}\ \bibnamefont {Peng}},
  \bibinfo {author} {\bibfnamefont {F.}~\bibnamefont {Nori}},\ and\ \bibinfo
  {author} {\bibfnamefont {Y.-x.}\ \bibnamefont {Liu}},\ }\bibfield  {title}
  {\bibinfo {title} {Topologically protected quantum coherence in a
  superatom},\ }\href {https://doi.org/10.1103/PhysRevLett.124.023603}
  {\bibfield  {journal} {\bibinfo  {journal} {Phys. Rev. Lett.}\ }\textbf
  {\bibinfo {volume} {124}},\ \bibinfo {pages} {023603} (\bibinfo {year}
  {2020})}\BibitemShut {NoStop}%
\bibitem [{\citenamefont {McDonnell}\ and\ \citenamefont
  {Olmos}(2022)}]{McDonnell2022subradiantedge}%
  \BibitemOpen
  \bibfield  {author} {\bibinfo {author} {\bibfnamefont {C.}~\bibnamefont
  {McDonnell}}\ and\ \bibinfo {author} {\bibfnamefont {B.}~\bibnamefont
  {Olmos}},\ }\bibfield  {title} {\bibinfo {title} {Subradiant edge states in
  an atom chain with waveguide-mediated hopping},\ }\href
  {https://doi.org/10.22331/q-2022-09-15-805} {\bibfield  {journal} {\bibinfo
  {journal} {{Quantum}}\ }\textbf {\bibinfo {volume} {6}},\ \bibinfo {pages}
  {805} (\bibinfo {year} {2022})}\BibitemShut {NoStop}%
\bibitem [{\citenamefont {Nie}\ \emph {et~al.}(2021)\citenamefont {Nie},
  \citenamefont {Antezza}, \citenamefont {Liu},\ and\ \citenamefont
  {Nori}}]{nie2021dissipative}%
  \BibitemOpen
  \bibfield  {author} {\bibinfo {author} {\bibfnamefont {W.}~\bibnamefont
  {Nie}}, \bibinfo {author} {\bibfnamefont {M.}~\bibnamefont {Antezza}},
  \bibinfo {author} {\bibfnamefont {Y.-x.}\ \bibnamefont {Liu}},\ and\ \bibinfo
  {author} {\bibfnamefont {F.}~\bibnamefont {Nori}},\ }\bibfield  {title}
  {\bibinfo {title} {Dissipative topological phase transition with strong
  system-environment coupling},\ }\href
  {https://doi.org/10.1103/PhysRevLett.127.250402} {\bibfield  {journal}
  {\bibinfo  {journal} {Phys. Rev. Lett.}\ }\textbf {\bibinfo {volume} {127}},\
  \bibinfo {pages} {250402} (\bibinfo {year} {2021})}\BibitemShut {NoStop}%
\bibitem [{\citenamefont {Lehmberg}(1970)}]{lehmberg1970radiation}%
  \BibitemOpen
  \bibfield  {author} {\bibinfo {author} {\bibfnamefont {R.~H.}\ \bibnamefont
  {Lehmberg}},\ }\bibfield  {title} {\bibinfo {title} {Radiation from an
  $n$-atom system. i. general formalism},\ }\href
  {https://doi.org/10.1103/PhysRevA.2.883} {\bibfield  {journal} {\bibinfo
  {journal} {Phys. Rev. A}\ }\textbf {\bibinfo {volume} {2}},\ \bibinfo {pages}
  {883} (\bibinfo {year} {1970})}\BibitemShut {NoStop}%
\bibitem [{\citenamefont {Dung}\ \emph {et~al.}(2002)\citenamefont {Dung},
  \citenamefont {Kn\"oll},\ and\ \citenamefont {Welsch}}]{dung2002resonant}%
  \BibitemOpen
  \bibfield  {author} {\bibinfo {author} {\bibfnamefont {H.~T.}\ \bibnamefont
  {Dung}}, \bibinfo {author} {\bibfnamefont {L.}~\bibnamefont {Kn\"oll}},\ and\
  \bibinfo {author} {\bibfnamefont {D.-G.}\ \bibnamefont {Welsch}},\ }\bibfield
   {title} {\bibinfo {title} {Resonant dipole-dipole interaction in the
  presence of dispersing and absorbing surroundings},\ }\href
  {https://doi.org/10.1103/PhysRevA.66.063810} {\bibfield  {journal} {\bibinfo
  {journal} {Phys. Rev. A}\ }\textbf {\bibinfo {volume} {66}},\ \bibinfo
  {pages} {063810} (\bibinfo {year} {2002})}\BibitemShut {NoStop}%
\bibitem [{\citenamefont {Ficek}\ and\ \citenamefont
  {Tanaś}(2002)}]{ficek2002entangled}%
  \BibitemOpen
  \bibfield  {author} {\bibinfo {author} {\bibfnamefont {Z.}~\bibnamefont
  {Ficek}}\ and\ \bibinfo {author} {\bibfnamefont {R.}~\bibnamefont {Tanaś}},\
  }\bibfield  {title} {\bibinfo {title} {Entangled states and collective
  nonclassical effects in two-atom systems},\ }\href
  {https://doi.org/https://doi.org/10.1016/S0370-1573(02)00368-X} {\bibfield
  {journal} {\bibinfo  {journal} {Physics Reports}\ }\textbf {\bibinfo {volume}
  {372}},\ \bibinfo {pages} {369} (\bibinfo {year} {2002})}\BibitemShut
  {NoStop}%
\bibitem [{\citenamefont {Le~Kien}\ \emph {et~al.}(2005)\citenamefont
  {Le~Kien}, \citenamefont {DuttaGupta}, \citenamefont {Nayak},\ and\
  \citenamefont {Hakuta}}]{le2005nanofiber}%
  \BibitemOpen
  \bibfield  {author} {\bibinfo {author} {\bibfnamefont {F.}~\bibnamefont
  {Le~Kien}}, \bibinfo {author} {\bibfnamefont {S.}~\bibnamefont {DuttaGupta}},
  \bibinfo {author} {\bibfnamefont {K.~P.}\ \bibnamefont {Nayak}},\ and\
  \bibinfo {author} {\bibfnamefont {K.}~\bibnamefont {Hakuta}},\ }\bibfield
  {title} {\bibinfo {title} {Nanofiber-mediated radiative transfer between two
  distant atoms},\ }\href {https://doi.org/10.1103/PhysRevA.72.063815}
  {\bibfield  {journal} {\bibinfo  {journal} {Phys. Rev. A}\ }\textbf {\bibinfo
  {volume} {72}},\ \bibinfo {pages} {063815} (\bibinfo {year}
  {2005})}\BibitemShut {NoStop}%
\bibitem [{\citenamefont {Gonzalez-Tudela}\ \emph {et~al.}(2011)\citenamefont
  {Gonzalez-Tudela}, \citenamefont {Martin-Cano}, \citenamefont {Moreno},
  \citenamefont {Martin-Moreno}, \citenamefont {Tejedor},\ and\ \citenamefont
  {Garcia-Vidal}}]{gonzalez2011entanglement}%
  \BibitemOpen
  \bibfield  {author} {\bibinfo {author} {\bibfnamefont {A.}~\bibnamefont
  {Gonzalez-Tudela}}, \bibinfo {author} {\bibfnamefont {D.}~\bibnamefont
  {Martin-Cano}}, \bibinfo {author} {\bibfnamefont {E.}~\bibnamefont {Moreno}},
  \bibinfo {author} {\bibfnamefont {L.}~\bibnamefont {Martin-Moreno}}, \bibinfo
  {author} {\bibfnamefont {C.}~\bibnamefont {Tejedor}},\ and\ \bibinfo {author}
  {\bibfnamefont {F.~J.}\ \bibnamefont {Garcia-Vidal}},\ }\bibfield  {title}
  {\bibinfo {title} {Entanglement of two qubits mediated by one-dimensional
  plasmonic waveguides},\ }\href
  {https://doi.org/10.1103/PhysRevLett.106.020501} {\bibfield  {journal}
  {\bibinfo  {journal} {Phys. Rev. Lett.}\ }\textbf {\bibinfo {volume} {106}},\
  \bibinfo {pages} {020501} (\bibinfo {year} {2011})}\BibitemShut {NoStop}%
\bibitem [{\citenamefont {Chang}\ \emph {et~al.}(2012)\citenamefont {Chang},
  \citenamefont {Jiang}, \citenamefont {Gorshkov},\ and\ \citenamefont
  {Kimble}}]{chang2012cavity}%
  \BibitemOpen
  \bibfield  {author} {\bibinfo {author} {\bibfnamefont {D.~E.}\ \bibnamefont
  {Chang}}, \bibinfo {author} {\bibfnamefont {L.}~\bibnamefont {Jiang}},
  \bibinfo {author} {\bibfnamefont {A.~V.}\ \bibnamefont {Gorshkov}},\ and\
  \bibinfo {author} {\bibfnamefont {H.~J.}\ \bibnamefont {Kimble}},\ }\bibfield
   {title} {\bibinfo {title} {Cavity {QED} with atomic mirrors},\ }\href
  {https://doi.org/10.1088/1367-2630/14/6/063003} {\bibfield  {journal}
  {\bibinfo  {journal} {New Journal of Physics}\ }\textbf {\bibinfo {volume}
  {14}},\ \bibinfo {pages} {063003} (\bibinfo {year} {2012})}\BibitemShut
  {NoStop}%
\bibitem [{\citenamefont {Lalumi\`ere}\ \emph {et~al.}(2013)\citenamefont
  {Lalumi\`ere}, \citenamefont {Sanders}, \citenamefont {van Loo},
  \citenamefont {Fedorov}, \citenamefont {Wallraff},\ and\ \citenamefont
  {Blais}}]{lalumiere2013input}%
  \BibitemOpen
  \bibfield  {author} {\bibinfo {author} {\bibfnamefont {K.}~\bibnamefont
  {Lalumi\`ere}}, \bibinfo {author} {\bibfnamefont {B.~C.}\ \bibnamefont
  {Sanders}}, \bibinfo {author} {\bibfnamefont {A.~F.}\ \bibnamefont {van
  Loo}}, \bibinfo {author} {\bibfnamefont {A.}~\bibnamefont {Fedorov}},
  \bibinfo {author} {\bibfnamefont {A.}~\bibnamefont {Wallraff}},\ and\
  \bibinfo {author} {\bibfnamefont {A.}~\bibnamefont {Blais}},\ }\bibfield
  {title} {\bibinfo {title} {Input-output theory for waveguide {QED} with an
  ensemble of inhomogeneous atoms},\ }\href
  {https://doi.org/10.1103/PhysRevA.88.043806} {\bibfield  {journal} {\bibinfo
  {journal} {Phys. Rev. A}\ }\textbf {\bibinfo {volume} {88}},\ \bibinfo
  {pages} {043806} (\bibinfo {year} {2013})}\BibitemShut {NoStop}%
\bibitem [{\citenamefont {Doyeux}\ \emph {et~al.}(2017)\citenamefont {Doyeux},
  \citenamefont {Gangaraj}, \citenamefont {Hanson},\ and\ \citenamefont
  {Antezza}}]{doyeux2017giant}%
  \BibitemOpen
  \bibfield  {author} {\bibinfo {author} {\bibfnamefont {P.}~\bibnamefont
  {Doyeux}}, \bibinfo {author} {\bibfnamefont {S.~A.~H.}\ \bibnamefont
  {Gangaraj}}, \bibinfo {author} {\bibfnamefont {G.~W.}\ \bibnamefont
  {Hanson}},\ and\ \bibinfo {author} {\bibfnamefont {M.}~\bibnamefont
  {Antezza}},\ }\bibfield  {title} {\bibinfo {title} {Giant interatomic
  energy-transport amplification with nonreciprocal photonic topological
  insulators},\ }\href {https://doi.org/10.1103/PhysRevLett.119.173901}
  {\bibfield  {journal} {\bibinfo  {journal} {Phys. Rev. Lett.}\ }\textbf
  {\bibinfo {volume} {119}},\ \bibinfo {pages} {173901} (\bibinfo {year}
  {2017})}\BibitemShut {NoStop}%
\bibitem [{\citenamefont {Mirhosseini}\ \emph {et~al.}(2019)\citenamefont
  {Mirhosseini}, \citenamefont {Kim}, \citenamefont {Zhang}, \citenamefont
  {Sipahigil}, \citenamefont {Dieterle}, \citenamefont {Keller}, \citenamefont
  {Asenjo-Garcia}, \citenamefont {Chang},\ and\ \citenamefont
  {Painter}}]{mirhosseini2019cavity}%
  \BibitemOpen
  \bibfield  {author} {\bibinfo {author} {\bibfnamefont {M.}~\bibnamefont
  {Mirhosseini}}, \bibinfo {author} {\bibfnamefont {E.}~\bibnamefont {Kim}},
  \bibinfo {author} {\bibfnamefont {X.}~\bibnamefont {Zhang}}, \bibinfo
  {author} {\bibfnamefont {A.}~\bibnamefont {Sipahigil}}, \bibinfo {author}
  {\bibfnamefont {P.~B.}\ \bibnamefont {Dieterle}}, \bibinfo {author}
  {\bibfnamefont {A.~J.}\ \bibnamefont {Keller}}, \bibinfo {author}
  {\bibfnamefont {A.}~\bibnamefont {Asenjo-Garcia}}, \bibinfo {author}
  {\bibfnamefont {D.~E.}\ \bibnamefont {Chang}},\ and\ \bibinfo {author}
  {\bibfnamefont {O.}~\bibnamefont {Painter}},\ }\bibfield  {title} {\bibinfo
  {title} {Cavity quantum electrodynamics with atom-like mirrors},\ }\href
  {https://doi.org/10.1038/s41586-019-1196-1} {\bibfield  {journal} {\bibinfo
  {journal} {Nature}\ }\textbf {\bibinfo {volume} {569}},\ \bibinfo {pages}
  {692} (\bibinfo {year} {2019})}\BibitemShut {NoStop}%
\bibitem [{\citenamefont {Wen}\ \emph {et~al.}(2019)\citenamefont {Wen},
  \citenamefont {Lin}, \citenamefont {Kockum}, \citenamefont {Suri},
  \citenamefont {Ian}, \citenamefont {Chen}, \citenamefont {Mao}, \citenamefont
  {Chiu}, \citenamefont {Delsing}, \citenamefont {Nori}, \citenamefont {Lin},\
  and\ \citenamefont {Hoi}}]{wen2019large}%
  \BibitemOpen
  \bibfield  {author} {\bibinfo {author} {\bibfnamefont {P.~Y.}\ \bibnamefont
  {Wen}}, \bibinfo {author} {\bibfnamefont {K.-T.}\ \bibnamefont {Lin}},
  \bibinfo {author} {\bibfnamefont {A.~F.}\ \bibnamefont {Kockum}}, \bibinfo
  {author} {\bibfnamefont {B.}~\bibnamefont {Suri}}, \bibinfo {author}
  {\bibfnamefont {H.}~\bibnamefont {Ian}}, \bibinfo {author} {\bibfnamefont
  {J.~C.}\ \bibnamefont {Chen}}, \bibinfo {author} {\bibfnamefont {S.~Y.}\
  \bibnamefont {Mao}}, \bibinfo {author} {\bibfnamefont {C.~C.}\ \bibnamefont
  {Chiu}}, \bibinfo {author} {\bibfnamefont {P.}~\bibnamefont {Delsing}},
  \bibinfo {author} {\bibfnamefont {F.}~\bibnamefont {Nori}}, \bibinfo {author}
  {\bibfnamefont {G.-D.}\ \bibnamefont {Lin}},\ and\ \bibinfo {author}
  {\bibfnamefont {I.-C.}\ \bibnamefont {Hoi}},\ }\bibfield  {title} {\bibinfo
  {title} {Large collective {Lamb} shift of two distant superconducting
  artificial atoms},\ }\href {https://doi.org/10.1103/PhysRevLett.123.233602}
  {\bibfield  {journal} {\bibinfo  {journal} {Phys. Rev. Lett.}\ }\textbf
  {\bibinfo {volume} {123}},\ \bibinfo {pages} {233602} (\bibinfo {year}
  {2019})}\BibitemShut {NoStop}%
\bibitem [{\citenamefont {van Loo}\ \emph {et~al.}(2013)\citenamefont {van
  Loo}, \citenamefont {Fedorov}, \citenamefont {Lalumière}, \citenamefont
  {Sanders}, \citenamefont {Blais},\ and\ \citenamefont
  {Wallraff}}]{van2013photon}%
  \BibitemOpen
  \bibfield  {author} {\bibinfo {author} {\bibfnamefont {A.~F.}\ \bibnamefont
  {van Loo}}, \bibinfo {author} {\bibfnamefont {A.}~\bibnamefont {Fedorov}},
  \bibinfo {author} {\bibfnamefont {K.}~\bibnamefont {Lalumière}}, \bibinfo
  {author} {\bibfnamefont {B.~C.}\ \bibnamefont {Sanders}}, \bibinfo {author}
  {\bibfnamefont {A.}~\bibnamefont {Blais}},\ and\ \bibinfo {author}
  {\bibfnamefont {A.}~\bibnamefont {Wallraff}},\ }\bibfield  {title} {\bibinfo
  {title} {Photon-mediated interactions between distant artificial atoms},\
  }\href {https://doi.org/10.1126/science.1244324} {\bibfield  {journal}
  {\bibinfo  {journal} {Science}\ }\textbf {\bibinfo {volume} {342}},\ \bibinfo
  {pages} {1494} (\bibinfo {year} {2013})}\BibitemShut {NoStop}%
\bibitem [{\citenamefont {Asenjo-Garcia}\ \emph {et~al.}(2017)\citenamefont
  {Asenjo-Garcia}, \citenamefont {Hood}, \citenamefont {Chang},\ and\
  \citenamefont {Kimble}}]{asenjo2017atom}%
  \BibitemOpen
  \bibfield  {author} {\bibinfo {author} {\bibfnamefont {A.}~\bibnamefont
  {Asenjo-Garcia}}, \bibinfo {author} {\bibfnamefont {J.~D.}\ \bibnamefont
  {Hood}}, \bibinfo {author} {\bibfnamefont {D.~E.}\ \bibnamefont {Chang}},\
  and\ \bibinfo {author} {\bibfnamefont {H.~J.}\ \bibnamefont {Kimble}},\
  }\bibfield  {title} {\bibinfo {title} {Atom-light interactions in
  quasi-one-dimensional nanostructures: A {Green's}-function perspective},\
  }\href {https://doi.org/10.1103/PhysRevA.95.033818} {\bibfield  {journal}
  {\bibinfo  {journal} {Phys. Rev. A}\ }\textbf {\bibinfo {volume} {95}},\
  \bibinfo {pages} {033818} (\bibinfo {year} {2017})}\BibitemShut {NoStop}%
\bibitem [{\citenamefont {Zhang}\ and\ \citenamefont
  {M\o{}lmer}(2019)}]{zhang2019theory}%
  \BibitemOpen
  \bibfield  {author} {\bibinfo {author} {\bibfnamefont {Y.-X.}\ \bibnamefont
  {Zhang}}\ and\ \bibinfo {author} {\bibfnamefont {K.}~\bibnamefont
  {M\o{}lmer}},\ }\bibfield  {title} {\bibinfo {title} {Theory of subradiant
  states of a one-dimensional two-level atom chain},\ }\href
  {https://doi.org/10.1103/PhysRevLett.122.203605} {\bibfield  {journal}
  {\bibinfo  {journal} {Phys. Rev. Lett.}\ }\textbf {\bibinfo {volume} {122}},\
  \bibinfo {pages} {203605} (\bibinfo {year} {2019})}\BibitemShut {NoStop}%
\bibitem [{\citenamefont {Zanner}\ \emph {et~al.}(2022)\citenamefont {Zanner},
  \citenamefont {Orell}, \citenamefont {Schneider}, \citenamefont {Albert},
  \citenamefont {Oleschko}, \citenamefont {Juan}, \citenamefont {Silveri},\
  and\ \citenamefont {Kirchmair}}]{zanner2022coherent}%
  \BibitemOpen
  \bibfield  {author} {\bibinfo {author} {\bibfnamefont {M.}~\bibnamefont
  {Zanner}}, \bibinfo {author} {\bibfnamefont {T.}~\bibnamefont {Orell}},
  \bibinfo {author} {\bibfnamefont {C.~M.~F.}\ \bibnamefont {Schneider}},
  \bibinfo {author} {\bibfnamefont {R.}~\bibnamefont {Albert}}, \bibinfo
  {author} {\bibfnamefont {S.}~\bibnamefont {Oleschko}}, \bibinfo {author}
  {\bibfnamefont {M.~L.}\ \bibnamefont {Juan}}, \bibinfo {author}
  {\bibfnamefont {M.}~\bibnamefont {Silveri}},\ and\ \bibinfo {author}
  {\bibfnamefont {G.}~\bibnamefont {Kirchmair}},\ }\bibfield  {title} {\bibinfo
  {title} {Coherent control of a multi-qubit dark state in waveguide quantum
  electrodynamics},\ }\href {https://doi.org/10.1038/s41567-022-01527-w}
  {\bibfield  {journal} {\bibinfo  {journal} {Nature Physics}\ }\textbf
  {\bibinfo {volume} {18}},\ \bibinfo {pages} {538} (\bibinfo {year}
  {2022})}\BibitemShut {NoStop}%
\end{thebibliography}%


\end{document}